\documentclass[aps,prc,twocolumn,superscriptaddress,showkeys,showpacs]{revtex4-2}
\usepackage{CJKutf8}
\usepackage{amssymb}
\usepackage{amsmath}
\usepackage{diagbox}
\usepackage{makecell}
\usepackage{tikz}
\usepackage{color}
\usepackage{graphicx} 
\usepackage{multirow} 
\usepackage{makecell}
\usepackage[normalem]{ulem}
\usepackage[hidelinks]{hyperref}
\usepackage{rotating} 
\usepackage{threeparttable} 
\usepackage{booktabs} 
\usepackage{longtable}  
\usepackage{array} 
\usepackage{bm} 
\usepackage{supertabular}
\usepackage{array} 

\newcommand{\red}[1]{#1}

\renewcommand{\arraystretch}{1.5} 
\begin{document}

\title{The $^{229}$Th Isomer: Nuclear Structure, Clocks, and Tests of Fundamental Physics}
 
\author{Xiao Lu}  
\email[]{luxiao@itp.ac.cn}
\affiliation{Institute of Theoretical Physics, Chinese Academy of Sciences, Beijing 100190, China}
\author{Rui Zhao}
\affiliation{Institute of Theoretical Physics, Chinese Academy of Sciences, Beijing 100190, China}
\affiliation{School of Physical Sciences, University of Chinese Academy of Sciences, Beijing 100049, China} 
\author{Shan-Gui Zhou} 
\email[]{sgzhou@itp.ac.cn}
\affiliation{Institute of Theoretical Physics, Chinese Academy of Sciences, Beijing 100190, China} 
\affiliation{School of Physical Sciences, University of Chinese Academy of Sciences, Beijing 100049, China} 
\affiliation{School of Nuclear Science and Technology, University of Chinese Academy of Sciences, Beijing 100049, China}  

\date{\today}

\begin{abstract}
The $^{229}$Th nucleus possesses an isomeric state at an excitation energy of $\sim 8$ eV, the lowest known nuclear transition energy, placing its frequency in the vacuum-ultraviolet range and making it directly accessible to laser spectroscopy. In this review, we discuss the $^{229}$Th isomer from three connected perspectives: experimental spectroscopy and clock development, nuclear structure theory, and applications to precision tests of fundamental physics. We first trace the experimental progress from indirect $\gamma$-ray energy inference to resonant laser excitation, absolute frequency comparison with an atomic clock, and feedback-loop operation of a solid-state nuclear clock, and discuss trapped-ion, highly charged ion, and solid-state platforms together with mechanisms for nuclear-state manipulation and readout. We then review, from the nuclear-structure perspective, how the near-degeneracy of the $5/2^+[633]$ and $3/2^+[631]$ neutron Nilsson configurations, together with Coriolis mixing and octupole correlations, underlies the anomalously low transition energy and its electromagnetic properties. Comparisons among different phenomenological and microscopic models show that octupole correlations are a common structural ingredient, while magnetic moments and transition strengths remain sensitive tests of the calculated wave functions. Finally, we discuss how the near-cancellation of MeV-scale nuclear contributions into an eV-scale transition can enhance sensitivity to variations of fundamental constants, signatures of ultralight dark matter, CP-violating interactions, Lorentz-invariance violation, and possible nuclear quantum technologies.
\end{abstract} 


\maketitle

\noindent\textbf{List of Abbreviations}
\par\smallskip

\begingroup
\footnotesize
\setlength{\tabcolsep}{4pt}
\renewcommand{\arraystretch}{1.0}
\begin{supertabular}{@{}>{\raggedright\arraybackslash}p{0.25\columnwidth}
                        >{\raggedright\arraybackslash}p{0.71\columnwidth}@{}}
\toprule
Abbreviation & Definition \\
\midrule
BCS & Bardeen--Cooper--Schrieffer \\
\red{CDFT} & \red{covariant density functional theory} \\
\red{\mbox{CQOM-DSM}} &
\red{coherent quadrupole--octupole mode combined with the deformed shell model} \\
\red{DFT} & \red{density functional theory} \\
\red{EB} & \red{electronic bridge} \\
\red{EDM} & \red{electric dipole moment} \\
\red{EEPV} & \red{Einstein equivalence principle violation} \\
\red{HCI} & \red{highly charged ion} \\
\red{HEB} & \red{hyperfine electronic bridge} \\
\red{HFBCS} &
\red{Hartree--Fock plus Bardeen--Cooper--Schrieffer} \\
\red{IC} & \red{internal conversion} \\
\red{LLIV} & \red{local Lorentz invariance violation} \\
\red{\mbox{MDC-RMF}} &
\red{multidimensionally-constrained relativistic mean-field} \\
\red{\mbox{MR-CDFT}} &
\red{multi-reference covariant density functional theory} \\
\red{\mbox{MR-DFT}} &
\red{multi-reference density functional theory} \\
\red{NEEC} & \red{nuclear excitation by electron capture} \\
\red{NHM} & \red{nuclear hyperfine mixing} \\
\red{PSM} & \red{projected shell model} \\
\red{QCD} & \red{quantum chromodynamics} \\
\red{\mbox{RAT-PRM}} &
\red{reflection-asymmetric triaxial particle-rotor model} \\
\red{RDFT} & \red{relativistic density functional theory} \\
\red{ULDM} & \red{ultralight dark matter} \\
\red{VUV} & \red{vacuum ultraviolet} \\
\bottomrule
\end{supertabular}
\endgroup

\section{Introduction} 

Nuclear isomers are metastable excited states of atomic nuclei whose lifetimes are significantly longer than the femtosecond-to-picosecond time scales of typical prompt nuclear transitions~\cite{Dracoulis2016RepProgPhysReview,Walker2020PhysScr100Years}. Their long lifetimes arise when electromagnetic decay is hindered by nuclear-structure constraints, such as large spin changes, shape coexistence, or $K$ quantum number forbiddenness~\cite{Dracoulis2016RepProgPhysReview,Walker2020PhysScr100Years,Misch2021Symmetry}. Among the approximately 3400 known nuclides, more than 2600 isomeric states with half-lives $T_{1/2}\ge 10\,{\rm ns}$ have been identified~\cite{Kondev2021NUBASE,Garg2023AtDataNuclDataTablesAtlas}. Most known nuclear isomers lie in the keV--MeV excitation-energy range, with half-lives extending from $10\,{\rm ns}$ to about $10^{16}$ years~\cite{Aprahamian2005NatPhysIsomer,Walker2020PhysScr100Years}. Their long lifetimes and the energy stored in nuclear excited states have motivated applications ranging from nuclear energy storage and nuclear medicine to $\gamma$-ray lasers and tests of fundamental physics~\cite{Aprahamian2005NatPhysIsomer,Dracoulis2016RepProgPhysReview,Walker2020PhysScr100Years}.

Across the spectrum of known nuclear isomers, the $^{229}$Th isomeric state ($^{229\mathrm{m}}\mathrm{Th}$) stands out for its exceptionally low excitation energy of $\sim8$ eV~\cite{Seiferle2019NatureEnergy,Perera2025PRHostDependent}. This energy scale places the transition frequency in the vacuum ultraviolet (VUV) spectrum, making it the only known nuclear transition directly accessible with optical/VUV laser spectroscopy~\cite{Tkalya1992JETPLettExcitation,Peik2003EurophysLettNuclear}. Such optical accessibility provides an unprecedented opportunity to investigate nuclear structure with laser-spectroscopic precision~\cite{Sikorsky2020PhysRevLettMeasurement}. Furthermore, the compact nature of nuclei and the shielding by the electron cloud enhance their stability against external perturbations, making $^{229}$Th a promising platform for a nuclear clock with the potential to reach fractional frequency uncertainty below $10^{-19}$~\cite{Campbell2012PhysRevLettSingleIon,Peik2021QuantSciTechnolNuclear}.
 
Experimentally, the transition energy has been progressively refined over five decades, from indirect $\gamma$-ray spectroscopy~\cite{Kroger1976NuclPhysA,Beck2007phys.rev.lett.Energya} to direct laser excitation in crystal hosts~\cite{Tiedau2024physrevlettLaser,Elwell2024physrevlettLaser,Zhang2024NatureThF4} and the first frequency comparison with an atomic clock~\cite{Zhang2024natureFrequency}. These advances have enabled two principal clock architectures: trapped-ion systems~\cite{Peik2003EurophysLettNuclear,Campbell2012PhysRevLettSingleIon}, which aim for high systematic accuracy through isolation from environmental perturbations, and solid-state platforms~\cite{Rellergert2010PhysRevLettConstraining,vonderWense2017PhysRevLettDirect}, which exploit large nuclear ensembles for high stability~\cite{Kazakov2012Performance}. 
Techniques for active manipulation of the nuclear state, including electronic-bridge (EB) pathways~\cite{Tkalya1992JETPLettExcitation,Wang2024Isomeric} and radiation-induced quenching~\cite{Hiraki2024natcommunControlling,Schaden2025phys.rev.researchLaserinduced}, are being developed to enable faster clock cycling and improved measurement duty cycle. In 2026, two independent closed-loop solid-state nuclear clocks were demonstrated by stabilizing continuous-wave laser systems to the 148~nm $^{229}$Th absorption resonance in Th:CaF$_2$, with one system compared against a Yb$^+$ ion clock and the other demonstrating cross-crystal frequency reproducibility~\cite{DeCol2026Feedback,Huang2026NuclearClock}.
 
A central question from the nuclear structure perspective is the origin of this anomalously low transition energy and the electromagnetic properties of the isomeric state. In the Nilsson framework of deformed nuclei, the ground state ($5/2^+[633]$) and the isomeric state ($3/2^+[631]$) are associated with a near-degeneracy of two neutron orbitals~\cite{Mottelson1955Phys.Rev.Classification,Barci2003Phys.Rev.CNucleara}, each serving as the bandhead of a rotational band. Coriolis mixing between these two rotational bands generates a nonzero $M1$ matrix element, but by itself cannot account for the measured electromagnetic moments~\cite{Dykhne1998JetpLett.Matrix}. Octupole deformation, a collective feature characteristic of the actinide region~\cite{Butler1996RevModPhys}, mixes opposite-parity orbitals near the Fermi surface and plays a key role in reproducing the observed $M1$ transition properties~\cite{Minkov2017Phys.Rev.Lett.Reduced}. Several theoretical frameworks have been applied to the structure of the $5/2[633]$ ground state and the $3/2[631]$ isomeric state, as well as their electromagnetic observables, including \red{the coherent quadrupole--octupole mode combined with the deformed shell model (CQOM-DSM)}~\cite{Minkov2017Phys.Rev.Lett.Reduced,Minkov2019Phys.Rev.Lett.Theoretical}, the Projected Shell Model (PSM)~\cite{Chen2025Phys.Lett.BPSM}, and nuclear density functional theory (DFT)-based approaches ranging from Skyrme Hartree-Fock-BCS and Skyrme multi-reference DFT to multi-reference covariant DFT and relativistic DFT combined with a particle-rotor model~\cite{Minkov2024Phys.Rev.CHFBCS,Restrepo2026DFT,zhou2025microscopic,Wang2026CPB}. Across these approaches, octupole correlations emerge as a common structural ingredient underlying the near-degeneracy and electromagnetic properties of the $^{229}$Th isomer.

Beyond explaining the level structure and electromagnetic observables, a microscopic description of the  $^{229}$Th isomer is needed to quantify its sensitivity to fundamental constants and to physics beyond the Standard Model. The anomalously low transition energy results from a near-cancellation between MeV-scale Coulomb and strong-interaction contributions, which amplifies the sensitivity of the transition frequency to variations of the fine-structure constant $\alpha$ and hadronic parameters~\cite{Flambaum2006PhysRevLettEnhanced,Peik2021QuantSciTechnolNuclear}. 
This makes $^{229}$Th a promising probe of ultralight dark matter, especially in the hadronic sector~\cite{Fuchs2025PhysRevXSearching,Caputo2025PhysRevCSensitivity}, of $P$- and $T$-violating interactions through the nuclear Schiff moment~\cite{Flambaum2019PhysRevCEnhanced}, and of possible violations of local Lorentz invariance and the Einstein equivalence principle through differential nuclear energy shifts~\cite{Flambaum2016PhysRevLettEnhancing,Dzuba2025PhysRevAUsing}. The potentially enormous intrinsic quality factor, $Q=\nu/\Delta\nu\sim 10^{19}$--$10^{20}$, reflects the extremely narrow natural linewidth of the nuclear transition relative to its optical frequency, providing the spectroscopic sharpness required for precision nuclear metrology~\cite{Peik2021QuantSciTechnolNuclear}, nuclear quantum optics~\cite{Wang2024Isomeric,Wang2026HyperfineResolved,Yu2025ChinPhysLettHighlyChargedTh6} and motivating proposals for nuclear-level qubits based on trapped $^{229}$Th ions~\cite{Wang2025DirectNuclearLevelQubits}.

In this review,  we summarize recent progress on the $^{229}$Th isomer from three perspectives: experimental spectroscopy and clock development, nuclear structure theory, and applications to fundamental physics and quantum technologies. In Sec.~\ref{sec:exp}, we trace the experimental progress from indirect energy inference to resonant laser excitation, atomic clock frequency comparison, and closed-loop clock operation, and discuss the main clock platforms, including trapped ions, highly charged ions, and solid-state systems, together with the mechanisms for nuclear state manipulation and readout. In Sec.~\ref{sec:theory}, we discuss the nuclear structure of $^{229}$Th, covering the Nilsson classification, Coriolis mixing, the CQOM-DSM framework, and modern microscopic approaches including the PSM and DFT-based methods, followed by a systematic comparison of theoretical predictions with experiment.  
In Sec.~\ref{sec:newphysics}, we explore applications of $^{229}$Th to searches for fundamental-constant variation and ultralight dark matter, tests of $P$- and $T$-violating interactions, possible violations of local Lorentz invariance and the Einstein equivalence principle, and nuclear quantum information processing. A summary and outlook are provided in Sec.~\ref{SP}.

\section{Experimental Progress on the $^{229}$Th Isomer and Nuclear Clock} \label{sec:exp}

Precise determination of the $^{229\mathrm{m}}\mathrm{Th}$ isomeric transition energy is crucial for connecting nuclear spectroscopy and precision laser metrology, and underpins the development of high accuracy nuclear clocks. A key milestone was reached in 2024 with the first direct frequency comparison between the $^{229}$Th nuclear transition in Th:CaF$_2$ and an optical $^{87}$Sr atomic clock~\cite{Zhang2024natureFrequency}, providing a kHz level frequency anchor for the nuclear transition. This advance supports two main experimental directions for nuclear clock development: trapped-ion platforms, rooted in the nuclear laser spectroscopy proposal~\cite{Peik2003EurophysLettNuclear,Gan2026CPBLinearIonTrap}, and solid-state systems based on $^{229}$Th-doped wide-bandgap crystals such as CaF$_2$ and LiSrAlF$_6$~\cite{Rellergert2010PhysRevLettConstraining,Kazakov2012Performance,Zhao2026CPBThDopedMaterials}. Closed-loop operation has now been demonstrated in two independent solid-state Th:CaF$_2$ nuclear-clock systems by stabilizing continuous-wave 148~nm lasers to the nuclear absorption resonance~\cite{DeCol2026Feedback,Huang2026NuclearClock}. In trapped-ion systems, schemes based on the electronic bridge provide routes for hyperfine-resolved excitation~\cite{Wang2024Isomeric,Yu2026CPBTh3TransitionFrequencies}. In solid-state systems, X-ray- and laser-induced quenching are being explored as active population control mechanisms to shorten the measurement cycle and improve duty cycle~\cite{Hiraki2024natcommunControlling,Schaden2025phys.rev.researchLaserinduced}. In this section, we first review the evolution of transition energy measurements, and then discuss the main nuclear clock platforms and state control mechanisms.

\begin{figure*}[ht!]
  \centering
  \includegraphics[width=1\linewidth]{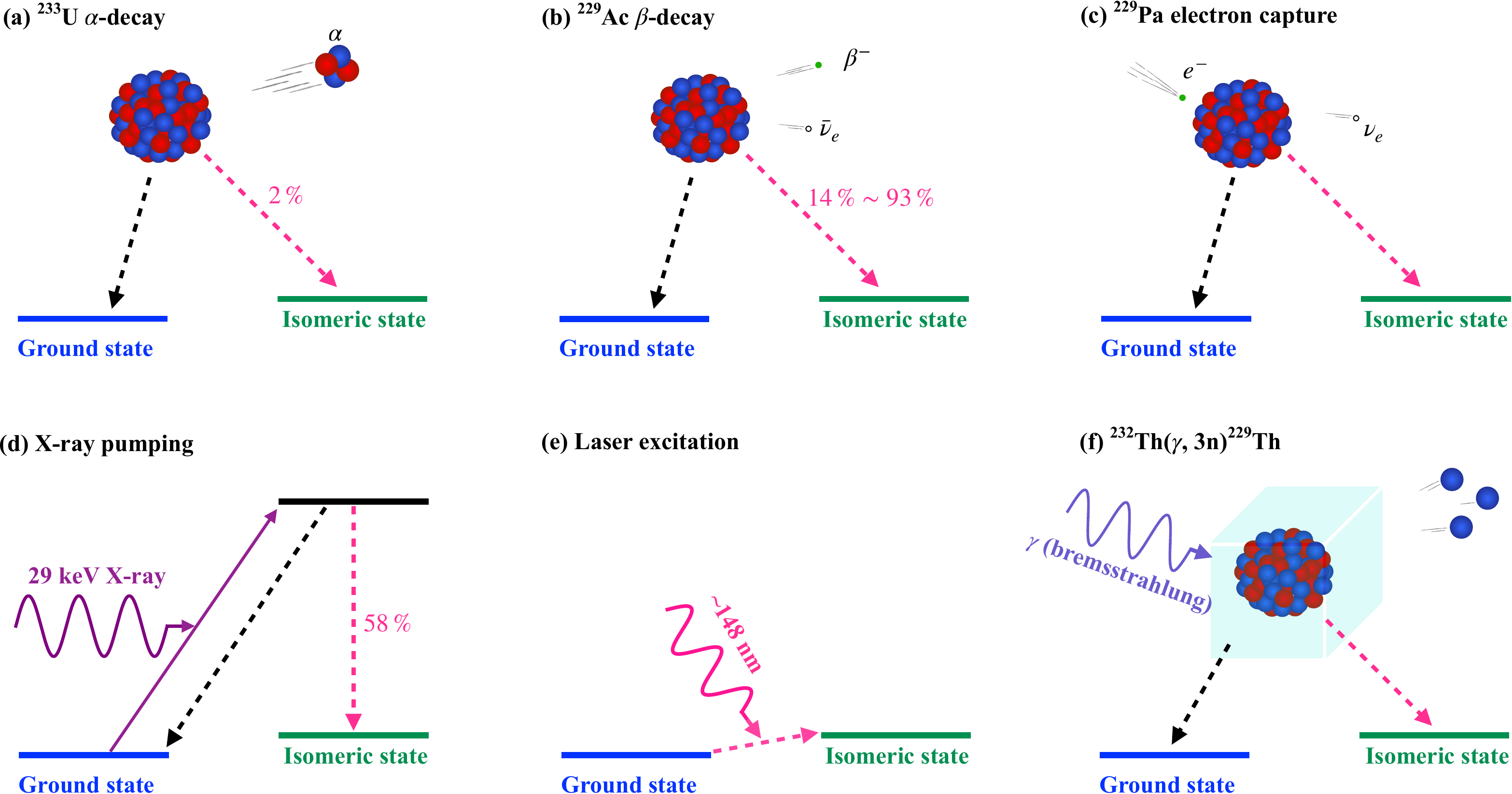}
  \caption{Schematic of representative pathways for populating the $^{229\mathrm{m}}\mathrm{Th}$
  isomeric state.
    (a) $\alpha$ decay of $^{233}$U~\cite{Kroger1976NuclPhysA,VonDerWense2016NatureDirecta};
    (b) $\beta$ decay of $^{229}$Ac~\cite{Ruchowska2006Phys.Rev.CNuclear};
    (c) possible electron-capture decay from $^{229}$Pa~\cite{Ahmad1982PhysRevLettPossible};
    (d) indirect X-ray pumping through the second excited state of $^{229}$Th at approximately 29~keV
    using resonant synchrotron radiation~\cite{Masuda2019natureXray};
    (e) direct resonant VUV laser
    excitation~\cite{Tiedau2024physrevlettLaser,Elwell2024physrevlettLaser};
    and (f) proposed photonuclear production of $^{229\mathrm{m},\mathrm{g}}\mathrm{Th}$ in Th-doped crystals via bremsstrahlung
    irradiation from electron accelerators~\cite{Lan2026AdvSciPhotonuclear}.}
      \label{fig:population_pathways}
\end{figure*}

A prerequisite for experimental studies of $^{229\mathrm{m}}\mathrm{Th}$ is the ability to populate the isomeric state. Existing approaches fall into three broad classes, as summarized in Fig.~\ref{fig:population_pathways}: population through radioactive decay or nuclear reactions, indirect pumping through higher nuclear levels, and direct resonant VUV excitation. Transfer reactions such as $^{230}$Th(d,t)$^{229}$Th have also provided spectroscopic information on the low-lying band structure~\cite{Burke1990phys.rev.cAdditional}. Radioactive-decay routes include the $\alpha$ decay of $^{233}$U with an approximately 2\% isomer branch~\cite{VonDerWense2016NatureDirecta,Yamaguchi2024natureLaser} and the $\beta$ decay of $^{229}$Ac with an estimated total feeding probability of 14\%--93\%~\cite{Ruchowska2006Phys.Rev.CNuclear,Kraemer2023natureObservation}; indirect pumping through the 29~keV second excited state, whose dominant in-band decay feeds the isomer~\cite{Masuda2019natureXray}; and direct resonant VUV excitation once the transition energy is sufficiently constrained~\cite{Tiedau2024physrevlettLaser,Elwell2024physrevlettLaser,Wang2026CPB148nmSources}. A recent variant of the nuclear-reaction route is photonuclear production in Th-doped crystals, which has been theoretically proposed and evaluated through Monte Carlo simulations as a route to generate $^{229\mathrm{m}, \mathrm{g}}\mathrm{Th}$ using bremsstrahlung irradiation, potentially bypassing the constrained $^{233}$U supply chain~\cite{Lan2026AdvSciPhotonuclear}.

\subsection{Energy Determination of the $^{229}\mathrm{Th}$ Isomer: From Indirect Measurement to Laser Excitation}

The first indirect indication of an exceptionally low-lying state in $^{229}$Th came from the $\alpha$-decay spectroscopy of $^{233}$U by Kroger and Reich, who inferred a $3/2^+$ band head lying within about $0.1~\mathrm{keV}$ of the ground state~\cite{Kroger1976NuclPhysA}.  Subsequent $\gamma$-ray energy difference measurements in the 1990s gave an eV-scale estimate, most notably $3.5\pm1.0~\mathrm{eV}$~\cite{Helmer1994phys.rev.cExciteda}, but were limited by systematic uncertainties. Motivated by this value, direct searches for optical or near-UV emission from the isomeric decay reported candidate signals~\cite{Irwin1997phys.rev.lett.Observation}; later reexaminations showed that the reported UV line structures were likely caused by $\alpha$-particle-induced fluorescence in air or N$_2$, rather than by the nuclear transition itself~\cite{Shaw1999PhysRevLettSpontaneous}. In 2007, X-ray microcalorimetry of $\gamma$ rays following $^{233}$U $\alpha$ decay revised the accepted transition energy to $7.6\pm0.5~\mathrm{eV}$~\cite{Beck2007phys.rev.lett.Energya}, moving experimental searches into the VUV range.

After the 2007 revision of the transition energy, experimental efforts shifted from indirect spectroscopic inference toward direct detection and active population of the isomer. In 2012, Zhao \textit{et al.} reported photon signals attributed to the radiative decay of $^{229\mathrm{m}}\mathrm{Th}$ with a half-life of $6\pm1$~h~\cite{Zhao2012phys.rev.lett.Observation}; however, this result has remained controversial because Peik and Zimmermann pointed out possible background contamination in the reported signal~\cite{Peik2013Comment}. The first unambiguous confirmation of the isomer came in 2016 through the detection of internal-conversion electrons emitted after neutralization of $^{229\mathrm{m}}\mathrm{Th}$~\cite{VonDerWense2016NatureDirecta}. Subsequent laser spectroscopy of $^{229\mathrm{m}}\mathrm{Th}^{2+}$ ions in 2018 determined the magnetic dipole and electric quadrupole moments of the isomer, as well as the change in mean-square charge radius~\cite{Thielking2018NatureLasera}. Between 2019 and 2020, high-precision studies based on internal-conversion electron spectroscopy~\cite{Seiferle2019NatureEnergy}, X-ray pumping via the 29-keV state~\cite{Masuda2019natureXray}, absolute $\gamma$-ray energy differences~\cite{Yamaguchi2019physrevlettEnergy}, and magnetic microcalorimetry~\cite{Sikorsky2020PhysRevLettMeasurement} constrained the transition energy to the VUV range relevant for wide-bandgap host crystals such as MgF$_2$ and CaF$_2$.

Entering the 2020s, the field progressed from identifying the isomeric signal to controlled excitation, decay manipulation, and clock-oriented frequency metrology. In 2023, the radiative decay of $^{229\mathrm{m}}\mathrm{Th}$ was observed in large-bandgap crystals~\cite{Kraemer2023natureObservation}. This result was followed in 2024 by resonant laser excitation of the nuclear transition in Th:CaF$_2$~\cite{Tiedau2024physrevlettLaser}, which was independently confirmed in Th:LiSrAlF$_6$~\cite{Elwell2024physrevlettLaser}. These advances enabled a direct frequency comparison between the $^{229}$Th nuclear transition and an $^{87}$Sr atomic clock using a VUV frequency comb, establishing the absolute transition frequency with high precision~\cite{Zhang2024natureFrequency}. In parallel, active control of the isomeric population has progressed through X-ray-induced quenching~\cite{Hiraki2024natcommunControlling} and laser-induced quenching~\cite{Schaden2025phys.rev.researchLaserinduced}, which provide mechanisms for depopulating the isomer and shortening the interrogation cycle in future solid-state clock operation. In 2026, closed-loop operation of solid-state $^{229}$Th nuclear-clock systems was independently demonstrated by locking continuous-wave 148~nm lasers to the nuclear absorption resonance in Th:CaF$_2$~\cite{DeCol2026Feedback,Huang2026NuclearClock}. Also in 2026, the measured half-life of singly charged $^{229\mathrm{m}}\mathrm{Th}^{+}$, $T_{1/2}=0.46(8)$~s, provided indirect evidence for EB s
decay~\cite{Shigekawa2026NaturePhysicsLifetime}.
 
\begin{figure*}[ht!]
  \centering
  \includegraphics[width=16 cm]{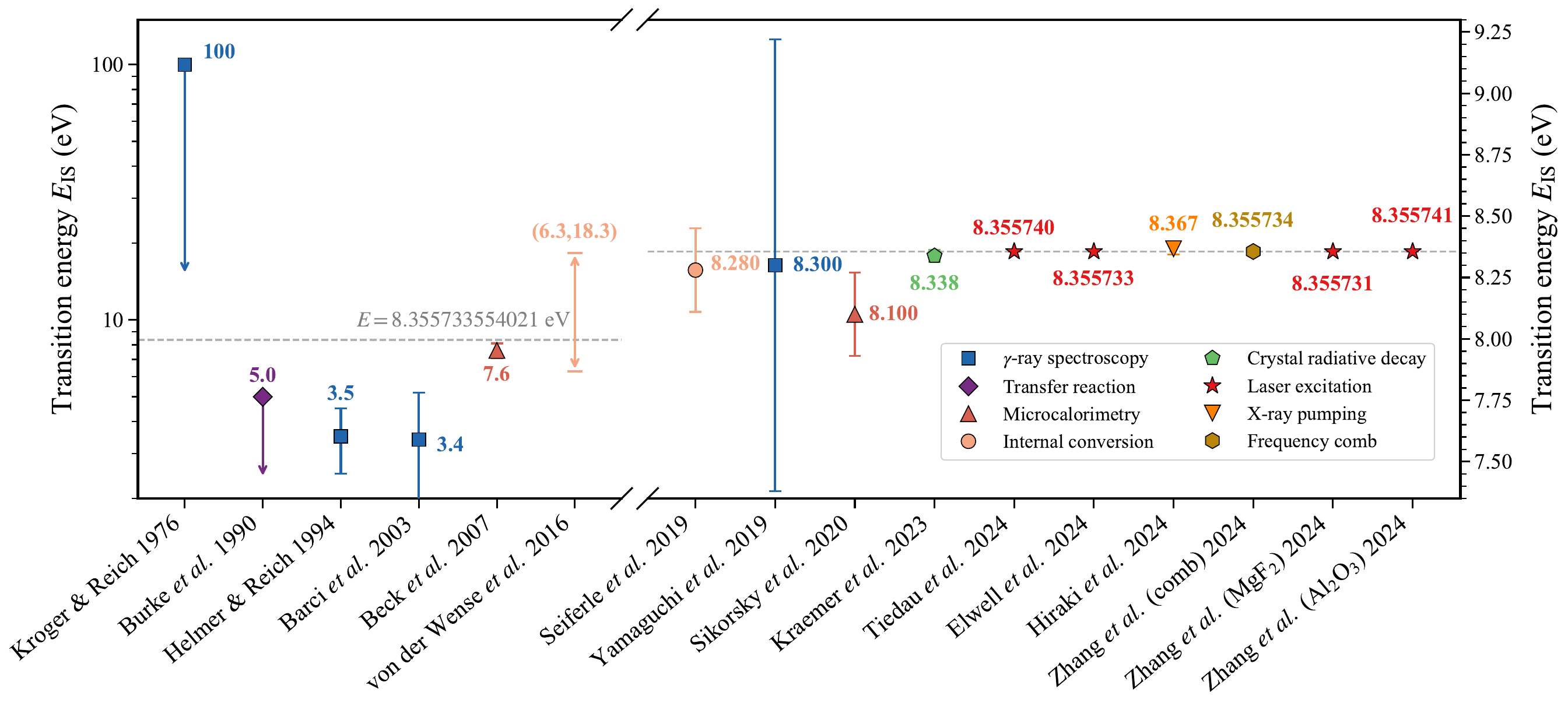}
  \caption{Historical evolution of the $^{229}$Th isomeric transition energy determinations and constraints. Measurements are arranged in chronological order within a broken-axis layout: the left panel shows early indirect constraints on a logarithmic energy scale, and the right panel zooms in on recent VUV-scale determinations. Downward arrows denote upper limits, vertical error bars denote quoted uncertainties, and the von der Wense \textit{et al.} interval $6.3<E_{\rm IS}<18.3$ eV is shown as a capped double-headed arrow. Marker shapes and colors indicate the experimental methods listed in the legend. The dashed horizontal line marks the current best value, $E_{\rm IS}=8.355\,733\,554\,021(8)$ eV, from VUV frequency-comb spectroscopy~\cite{Zhang2024natureFrequency}. Data are taken from the measurements summarized in Table~\ref{th229-exp}. The 1990 value $-1\pm4$ eV~\cite{Reich1990phys.rev.lett.Energya} is not plotted because it is nonpositive and cannot be represented on a logarithmic energy axis.}
  \label{fig:isomer_energy_history}
\end{figure*}

These experimental milestones are summarized in two complementary tables. Table~\ref{th229-exp} focuses on the determinations and constraints of the isomer energy $E_{\rm IS}$, whose progressive refinement over five decades is illustrated in Fig.~\ref{fig:isomer_energy_history}, whereas Table~\ref{th229-obs} collects nuclear observables relevant to decay and structure, including half-lives or lifetimes, magnetic dipole moments and ratios, electric quadrupole and hexadecapole moments and ratios, magnetic $g$-factor quantities, the reduced magnetic-dipole transition probability $B(M1)$, and the difference between the mean-square charge radii of the ground and isomeric states.

\begin{table*}[ht!]
    \centering
    \caption{Experimental determinations and constraints of the $^{229}$Th isomer energy or transition frequency. Early entries are indirect constraints from nuclear spectroscopy or transfer reactions, whereas later entries include internal-conversion spectroscopy, $\gamma$-ray and microcalorimetric determinations, radiative-decay measurements in crystals, X-ray pumping, and resonant VUV laser spectroscopy in solid-state hosts.}
    \label{th229-exp}
    \small
    \resizebox{\textwidth}{!}{%
    \begin{tabular}{llll}
        \hline\hline
        \textbf{Year} & \textbf{Method/system} & \textbf{$E_{\rm IS}$} & \textbf{Comment} \\
        \hline
        1976~\cite{Kroger1976NuclPhysA} & $^{233}$U $\alpha$ decay, $\gamma$-ray spectroscopy & $\lesssim 100$ eV & $3/2^+[631]$ bandhead inferred within $\sim 0.1$ keV of the ground state \\
        1990~\cite{Reich1990phys.rev.lett.Energya} & $^{233}$U $\alpha$ decay, $\gamma$-ray energy differences & $-1\pm4$ eV & Implies $E_{\rm IS}<10$ eV \\
        1990~\cite{Burke1990phys.rev.cAdditional} & $^{230}$Th$(d,t)^{229}$Th transfer reaction & $\leq5$ eV & Additional evidence for a near-ground-state level \\
        1994~\cite{Helmer1994phys.rev.cExciteda} & $^{233}$U $\alpha$ decay, improved $\gamma$-ray energy differences & $3.5\pm1.0$ eV & Historical benchmark before direct detection \\
        2003~\cite{Barci2003Phys.Rev.CNucleara} & $^{233}$U $\alpha$ decay, $\gamma$-ray spectroscopy & $3.4\pm1.8$ eV & Nuclear-structure reassessment of the low-energy bandhead \\
        2007~\cite{Beck2007phys.rev.lett.Energya} & $\gamma$-ray microcalorimetry & $7.6\pm0.5$ eV & Ground-state doublet splitting \\
        2016~\cite{VonDerWense2016NatureDirecta} & Direct detection of IC electrons & $6.3<E_{\rm IS}<18.3$ eV & Constraint from charge-state-dependent IC energetics \\
        2019~\cite{Seiferle2019NatureEnergy} & IC electron spectroscopy & $8.28\pm0.17$ eV & Direct energy measurement through IC decay \\
        2019~\cite{Yamaguchi2019physrevlettEnergy} & Absolute $\gamma$-ray energy difference & $8.30\pm0.92$ eV & Independent $\gamma$-spectroscopic determination \\
        2020~\cite{Sikorsky2020PhysRevLettMeasurement} & Magnetic microcalorimetry & $8.10(17)$ eV & Independent confirmation of the VUV energy range \\
        2023~\cite{Kraemer2023natureObservation} & Radiative decay in large-bandgap crystals & $8.338(24)$ eV & VUV photon emission from $^{229\mathrm{m}}$Th \\
        2024~\cite{Tiedau2024physrevlettLaser} & Laser excitation in Th:CaF$_2$ & $8.35574(3)$ eV & First resonant laser excitation in CaF$_2$ \\
        2024~\cite{Elwell2024physrevlettLaser} & Laser excitation in Th:LiSrAlF$_6$ & $8.355733(2)_{\rm stat}(10)_{\rm sys}$ eV & Independent solid-state confirmation \\
        2024~\cite{Hiraki2024natcommunControlling} & X-ray pumping in Th:CaF$_2$ & $8.367\pm0.024$ eV & Population and radiative decay in a VUV-transparent crystal \\
        2024~\cite{Zhang2024natureFrequency} & VUV frequency comb, comparison with $^{87}$Sr clock & $8.355733554021(8)$ eV & Frequency ratio measurement in CaF$_2$ at 150(1) K \\
        2024~\cite{Zhang2024NatureThF4} & Laser excitation in $^{229}$ThF$_4$ thin films & $8.355731(2)_{\rm stat}(124)_{\rm sys}$ eV; $8.355741(3)_{\rm stat}(124)_{\rm sys}$ eV & Converted from line-center frequencies measured for MgF$_2$ and Al$_2$O$_3$ substrates \\
        \hline\hline
    \end{tabular}
    }
\end{table*}

\begin{table*}[ht!]
    \centering
    \caption{Chronological overview of experimental nuclear observables for $^{229}$Th, excluding the energy values already listed in Table~\ref{th229-exp}. The table collects decay half-lives $T_{1/2}$ or lifetimes $\tau$, magnetic dipole moments $\mu_{\rm GS}$ and $\mu_{\rm IS}$ and their ratios, electric quadrupole and hexadecapole moments, including intrinsic moments $Q_{\lambda0}$, spectroscopic moments $Q_{\lambda}^{s}$, and IS/GS moment ratios, magnetic $g$-factor quantities such as $g_K$ and $g_K-g_R$, the reduced magnetic-dipole transition probability $B(M1)$, and the difference between the mean-square charge radii of the isomeric and ground states, $\delta\langle r^2\rangle=\langle r^2\rangle_{\rm IS}-\langle r^2\rangle_{\rm GS}$. 
    The subscripts GS and IS denote the ground and isomeric states, respectively. Intrinsic and spectroscopic multipole moments are denoted by $Q_{\lambda0}$ and $Q_{\lambda}^{s}$, respectively, with $\lambda=2$ and $4$ corresponding to quadrupole and hexadecapole moments.}
    \label{th229-obs}
    \small
    \resizebox{\textwidth}{!}{%
    \begin{tabular}{llll}
        \hline\hline
        \textbf{Year} & \textbf{Method/system} & \textbf{Observable} & \textbf{Reference} \\
        \hline
        1970 & $^{233}$U $\alpha$ decay, rotational-state lifetimes & $g_K=0.09\pm0.03$ for the ground-state rotational band & \cite{Ton1970nucl.phys.aLIFETIMES} \\
        1974 & Atomic hyperfine spectroscopy & $\mu_{\rm GS}=(+0.45\pm0.04)\,\mu_N$; $Q_{\rm GS}^{s}=+4.3\pm0.9\,e\,{\rm b}$ & \cite{Gerstenkorn1974} \\
        1988 & Coulomb excitation & $Q_{\rm GS}^{s}=3.149\pm0.032\,e\,{\rm b}$; $Q_{4,\rm GS}^{s}=0.088\pm0.017\,e\,{\rm b}^2$; $Q_{20}=8.816\pm0.090\,e\,{\rm b}$; $Q_{40}=3.69\pm0.72\,e\,{\rm b}^2$; $g_K-g_R=-(0.181\pm0.005)$ & \cite{Bemis1988Coulomb} \\
        2003 & $^{233}$U $\alpha$ decay, $\gamma$-ray spectroscopy & $Q_{20}=7.1\pm0.3\,e\,{\rm b}$; $|g_K-g_R|=0.176\pm0.021$ & \cite{Barci2003Phys.Rev.CNucleara} \\
        2011 & Laser-cooled $^{229}$Th$^{3+}$ Wigner crystals & $Q_{\rm GS}^{s}=3.11\pm0.16\,e\,{\rm b}$ & \cite{2011PhysRevLett.106.223001} \\
        2012 & Reported photon signal from $^{233}$U $\alpha$ decay & Reported half-life $\sim6\pm1$ h, disputed & \cite{Zhao2012phys.rev.lett.Observation,Peik2013Comment} \\
        2016 & Direct IC-electron detection & $T_{1/2}(^{229\mathrm{m}}\mathrm{Th}^{2+})>60$ s & \cite{VonDerWense2016NatureDirecta} \\
        2017 & Neutral Th, internal conversion decay & $T_{1/2}(\mathrm{IC})=7\pm1\,\mu$s & \cite{Seiferle2017Phys.Rev.Lett.Lifetime} \\
        2018 & Laser spectroscopy of trapped $^{229\mathrm{m}}$Th$^{2+}$ & $\mu_{\rm IS}/\mu_{\rm GS}=-1.04(15)$; $Q_{\rm IS}^{s}/Q_{\rm GS}^{s}=0.555(19)$; $\delta\langle r^2\rangle=0.012(2)$ fm$^2$ & \cite{Thielking2018NatureLasera} \\
        2021 & Radiative half-life estimate from excited-state lifetimes & $B(M1)=0.014(3)\,\mu_N^2$ ($0.0078(17)$ W.u.); $T_{1/2}(M1)=5.0(11)\times10^3$ s & \cite{Shigekawa2021phys.rev.cEstimation} \\
        2023 & Radiative decay in MgF$_2$ & $T_{1/2}=670(102)$ s in MgF$_2$; $T_{1/2}=2.21(34)\times10^3$ s in vacuum & \cite{Kraemer2023natureObservation} \\
        2024 & Laser spectroscopy of trapped $^{229\mathrm{m}}$Th$^{3+}$ & $T_{1/2}(^{229\mathrm{m}}\mathrm{Th}^{3+})=1400^{+600}_{-300}$ s; $\mu_{\rm IS}=-0.378(8)\,\mu_N$; $Q_{\rm IS}^{s}=1.77(2)\,e\,{\rm b}$ ($Q_{0,\rm IS}=8.84(10)\,e\,{\rm b}$); $\delta\langle r^2\rangle=0.0097(26)$ fm$^2$ & \cite{Yamaguchi2024natureLaser} \\
        2024 & Laser excitation in Th:CaF$_2$ & $\tau=630(15)$ s in CaF$_2$; $T_{1/2}=1740(50)$ s in vacuum; $B(M1)=0.022$ W.u. & \cite{Tiedau2024physrevlettLaser} \\
        2024 & Laser excitation in Th:LiSrAlF$_6$ & $\tau=568(13)_{\rm stat}(20)_{\rm sys}$ s in LiSrAlF$_6$ & \cite{Elwell2024physrevlettLaser} \\
        2024 & X-ray pumping in Th:CaF$_2$ & $T_{1/2}=447(25)$ s in CaF$_2$ & \cite{Hiraki2024natcommunControlling} \\
        2024 & VUV frequency comb spectroscopy in Th:CaF$_2$ & $Q_{\rm IS}^{s}/Q_{\rm GS}^{s}=0.57003(1)$; $\tau=641(4)$ s in CaF$_2$ & \cite{Zhang2024natureFrequency} \\
        2026 & Singly charged $^{229\mathrm{m}}$Th$^+$ & $T_{1/2}(^{229\mathrm{m}}\mathrm{Th}^{+})=0.46(8)$ s & \cite{Shigekawa2026NaturePhysicsLifetime} \\
        \hline\hline
        \multicolumn{4}{l}{\footnotesize Notes: Host-dependent lifetimes or decay constants should not be interpreted as bare-nucleus radiative lifetimes without the corresponding environmental correction. For the $M1$ strength, $1~{\rm W.u.}=1.7905\,\mu_N^2$.}
    \end{tabular}
    }
\end{table*}

\subsection{Nuclear Clock Platforms and Nuclear-State Control}

With the transition frequency now precisely identified, experimental efforts toward $^{229}$Th nuclear clocks have focused on two principal architectures: trapped-ion systems and solid-state systems, which emphasize high systematic accuracy and large-ensemble stability, respectively. Highly charged ions provide a complementary route for studying and exploiting nuclear-electronic coupling, including EB and nuclear-hyperfine-mixing effects, rather than a separate mature clock architecture.

\subsubsection{Trapped-Ion Nuclear Clocks}

Trapped-ion nuclear clocks represent one of the earliest detailed conceptual frameworks for a $^{229\mathrm{m}}\mathrm{Th}$ frequency standard, first proposed by Peik and Tamm in 2003~\cite{Peik2003EurophysLettNuclear}. This architecture uses either a single $^{229}\text{Th}$ ion or small ensembles of $^{229}\text{Th}$ ions confined in radio-frequency (Paul) traps under ultra-high vacuum~\cite{Peik2003EurophysLettNuclear,Berengut2009PhysRevLettProposed,Wense2020EurPhysJThe229Th,Gan2026CPBLinearIonTrap}. The $^{229}\text{Th}^{3+}$ charge state is typically selected for its one-valence-electron, alkali-like electronic structure. The operational principle relies on an electron-nuclear double-resonance scheme: a first laser is kept resonant with a closed electronic transition to produce continuous resonance fluorescence, while a second VUV laser interrogates the nuclear transition~\cite{Peik2003EurophysLettNuclear}. When nuclear excitation occurs, the resulting changes in nuclear spin and moments alter the electronic hyperfine structure, causing a detectable drop in fluorescence, thereby providing a state-selective signature of nuclear excitation. This largely non-destructive detection is uniquely enabled by the $^{229}\text{Th}^{3+}$ ion's electronic configuration, which simultaneously provides the closed transitions required for direct laser cooling and state preparation~\cite{Peik2003EurophysLettNuclear,2011PhysRevLett.106.223001,Thielking2018NatureLasera,Yamaguchi2024natureLaser,Yu2026CPBTh3TransitionFrequencies}.

To function as a frequency standard, the detection scheme is integrated into a closed-loop control system. The fluorescence signal serves as the discriminant for a feedback loop that continuously steers the VUV interrogation laser (or a tooth of a VUV frequency comb) to the nuclear resonance peak~\cite{Wense2020EurPhysJThe229Th}. Furthermore, to overcome the long radiative lifetime ($\sim 10^3$ s) that would otherwise limit the clock's sampling rate and stability, proposals based on the EB/hyperfine electronic bridge (HEB) mechanism explore active population transfer between the nuclear states~\cite{Wang2024Isomeric,Bilous2020PhysRevLettElectronic}, with the goal of shortening state-preparation or reset times in future clock cycles.

The primary advantage of the ion-trap platform is its potential for very high systematic accuracy, with a total fractional inaccuracy approaching $1\times10^{-19}$ estimated for a single-ion nuclear clock~\cite{Campbell2012PhysRevLettSingleIon}.
By isolating the nucleus from the external environment, perturbations such as blackbody radiation and Stark shifts are minimized, while the use of specific ``stretched'' hyperfine states allows for the suppression of external field-induced frequency shifts~\cite{Campbell2012PhysRevLettSingleIon}. 
Current progress is characterized by the advancements in trapping and cooling techniques. Several laboratories, including PTB~\cite{Thielking2018NatureLasera}, GIT~\cite{2011PhysRevLett.106.223001}, and RIKEN~\cite{Yamaguchi2024natureLaser}, have successfully achieved stable ion confinement. A landmark breakthrough occurred in 2024 when Yamaguchi \textit{et al.} performed high-resolution laser spectroscopy of the $^{229}\text{Th}^{3+}$ isomeric state in an ion-trap apparatus~\cite{Yamaguchi2024natureLaser}. Furthermore, by utilizing sympathetic cooling with $^{88}\text{Sr}^+$ to reach millikelvin temperatures, researchers have strongly reduced Doppler-related uncertainties~\cite{Zitzer2025PhysRevASympathetic}. Most recently, in 2026, the measurement of the $^{229\text{m}}\text{Th}^+$ half-life ($0.46 \pm 0.08$~s) indirectly suggested the existence of the EB decay mechanism~\cite{Shigekawa2026NaturePhysicsLifetime}, supporting the relevance of EB decay for nuclear-state depopulation and future active manipulation schemes. Theoretical proposals for hyperfine-resolved laser excitation and detection schemes~\cite{Wang2026HyperfineResolved} further outline pathways toward a fully operational trapped-ion nuclear clock. Consequently, the trapped-ion platform is expected to provide the highest spectroscopic resolution and lowest systematic uncertainty, establishing it as a primary reference for high-sensitivity tests of fundamental physical constants~\cite{Peik2021QuantSciTechnolNuclear}.

The ion-trap concept naturally extends to highly charged $^{229}\text{Th}$ ions, which are manipulated within electron beam ion traps or heavy-ion storage rings~\cite{Bilous2020PhysRevLettElectronic, Strizhov1991SovPhysJETPDecay, Ma2015PhysScrProposal}. In this regime, the electronic shell acts as a resonant antenna to bypass the nucleus's exceedingly small direct photo-excitation cross-section~\cite{Tkalya1992JETPLettExcitation}. In EB, laser-driven electronic excitation is coupled to the nucleus through virtual electronic states, whereas in NEEC a free electron is captured into a bound shell while transferring its excess energy to the nucleus~\cite{Tkalya1992JETPLettExcitation,Bilous2020PhysRevLettElectronic,Strizhov1991SovPhysJETPDecay}. In open-4f-shell $^{229}\text{Th}^{35+}$, the EB process has been proposed as an efficient tool for nuclear state manipulation~\cite{Bilous2020PhysRevLettElectronic}. For other highly charged ions, the HEB scheme offers new pathways for probing $\alpha$-variation~\cite{Wang2025PhysRevASearch}. A qualitatively different regime is reached in hydrogen-like $^{229}\text{Th}^{89+}$, where the remaining electron generates an ultra-strong magnetic field ($\sim 28$~MT) that strongly mixes the wavefunctions of the ground and isomeric states via nuclear hyperfine mixing (NHM)~\cite{Shabaev2022PhysRevLettDetermine}. This mixing breaks original selection rules, shortening the isomer lifetime from the radiative scale of hours to a few tens of milliseconds~\cite{Shabaev2022PhysRevLettDetermine}. While the dramatically shortened lifetime makes Th$^{89+}$ unsuitable as a clock oscillator, the NHM effect demonstrates a fundamental nuclear-electronic coupling phenomenon~\cite{PhysRevLett.133.152503} and offers a novel probe for nuclear structure~\cite{Wang2025PhysRevASearch}. Theoretical proposals suggest an alternative spectroscopic approach in these extreme environments through the indirect determination of nuclear transition properties. For instance, nuclear transition amplitudes can be extracted from precision measurements of the $g$-factor in single-electron $^{229}\text{Th}^{89+}$ ions~\cite{Shabaev2022PhysRevLettDetermine}.  More recently, a complementary HCI hyperfine-spectroscopy scheme using $^{229}\mathrm{Th}^{87+}$ and $^{229}\mathrm{Th}^{79+}$ was proposed to determine $\mu_{\rm GS}$, $\mu_{\rm IS}$, the bare nuclear transition energy, $\delta\langle r^2\rangle$, and the $M1$ transition matrix element within a single overdetermined framework~\cite{Zheng2026arXivHCINuclearParameters}.

\subsubsection{Solid-State Nuclear Clocks}

Research on solid-state nuclear clocks encompasses two primary architectures: the internal conversion (IC)-suppressed nuclear clock and the IC-based nuclear clock. 
The first detailed concept and metrological analysis of a nuclear clock based on the $^{229}$Th isomer was presented by Peik and Tamm in 2003~\cite{Peik2003EurophysLettNuclear}, while the IC-suppressed architecture was pioneered in 2010 by Rellergert \textit{et al.}~\cite{Rellergert2010PhysRevLettConstraining} and developed into a full performance analysis by Kazakov \textit{et al.} in 2012~\cite{Kazakov2012Performance}.    
This approach involves embedding $^{229}\text{Th}$ nuclei as dopants within wide-bandgap, VUV transparent crystals~\cite{Tiedau2024physrevlettLaser,Elwell2024physrevlettLaser} or developing thin-film architectures like $^{229}\text{ThF}_4$~\cite{Zhang2024NatureThF4}. 
In wide-bandgap hosts, the electronic gap can suppress ordinary IC and allow radiative decay to become observable or dominant at suitable lattice sites, while residual quenching and defect-related channels remain material-dependent issues~\cite{Dessovic2014JPhysCondensThorium,Tkalya2000JETPLettSpontaneous,Hiraki2024natcommunControlling,Xu2026CPBNucleusElectronEnvironment}. 

The defining advantage of this platform is its large number of quantum oscillators ($10^{15}$ to $10^{18} \text{ cm}^{-3}$), which provides an exceptionally high signal-to-noise ratio and superior short-term frequency stability, with the potential to reach fractional-instability levels around $10^{-19}$ under idealized operating conditions~\cite{Kazakov2012Performance,Hiraki2024natcommunControlling}. Recent years have seen rapid experimental advancements: following the 2023 observation of radiative decay in doped crystals~\cite{Kraemer2023natureObservation}, a key milestone was reached in 2024 with the first resonant laser excitation of $^{229}\text{Th}$ in $\text{CaF}_2$~\cite{Tiedau2024physrevlettLaser} and $\text{LiSrAlF}_6$~\cite{Elwell2024physrevlettLaser} host crystals, as well as in $^{229}\text{ThF}_4$ thin films~\cite{Zhang2024NatureThF4}. The $^{229}\text{ThF}_4$ architecture is notable for employing a stoichiometric thorium compound rather than a dilute doping scheme. \red{A subsequent study measured} the \red{temperature dependence of the four strongest} resolved quadrupole components~\cite{Higgins2025PhysRevLettTemperature}\red{.} Ooi \textit{et al.} \red{found that the first-order temperature dependence of} the $m=\pm5/2\to\pm3/2$ \red{transition} in $^{229}$Th:CaF$_2$ \red{vanishes at} $196(5)\,\mathrm{K}$~\cite{Ooi2026NatureFrequency}. 
In 2026, two independent feedback-loop solid-state $^{229}$Th nuclear-clock systems were reported. De Col \textit{et al.} stabilized a continuous-wave 148~nm laser to the nuclear absorption resonance in room-temperature Th:CaF$_2$ and continuously compared a subharmonic of the VUV light with a Yb$^+$ single-ion clock, reporting shot-noise-limited fractional-instability scaling of $3\times10^{-12}/\sqrt{\tau/\mathrm{s}}$ and one-day continuous operation~\cite{DeCol2026Feedback}. Independently, Huang \textit{et al.} used a cadmium-vapor resonance-enhanced four-wave-mixing source producing $\sim$10~$\mu$W of continuous-wave 148.4~nm radiation and phototube-based absorption readout, achieving $2\times10^{-12}/\sqrt{\tau/\mathrm{s}}$ instability and a cross-crystal frequency difference of 558(131)~Hz, corresponding to $2.8(0.6)\times10^{-13}$~\cite{Huang2026NuclearClock}. These \red{advances in temperature control and feedback} pave the way for portable, high-stability nuclear frequency references~\cite{Wang2026CPB148nmSources,Zhao2026CPBThDopedMaterials}.

\red{Despite this progress, clocks based on transparent host crystals remain subject to systematic frequency shifts and line broadening caused by interactions with the surrounding lattice. The electric field gradient at a thorium lattice site couples to the nuclear quadrupole moment, producing electric-quadrupole splitting and associated shifts of the individual transition frequencies~\cite{Kazakov2012Performance,Dessovic2014JPhysCondensThorium}. The difference in nuclear charge distribution between the ground and isomeric states gives rise to an isomer shift whose magnitude depends on the electronic contact density at the thorium site and therefore on the host material and lattice site~\cite{Perera2025PRHostDependent,Xu2026CPBNucleusElectronEnvironment}. Local charge-compensation configurations, lattice defects, dopant-induced strain, and inequivalent thorium sites can generate site-dependent frequency offsets and inhomogeneous broadening, while magnetic coupling to neighboring host nuclei provides an additional decoherence channel~\cite{Kazakov2012Performance,Dessovic2014JPhysCondensThorium,Ooi2026NatureFrequency,Xu2026CPBNucleusElectronEnvironment}. These effects influence both clock accuracy and stability. Host- and sample-dependent frequency shifts limit frequency reproducibility and can contribute to clock instability when they vary over time; line broadening and decoherence reduce spectroscopic contrast, degrade short-term stability, and complicate determination of the line center.} 

In contrast, the IC-based nuclear clock takes a different approach. Originally proposed in 2017~\cite{vonderWense2017PhysRevLettDirect} and refined into a comprehensive design in 2020~\cite{vonderWense2020EurPhysJDConcepts}, this scheme deposits $^{229}\text{Th}$ as thin metallic films or compounds such as $\text{ThO}_2$ on substrates. Rather than suppressing IC by nine orders of magnitude, it fully exploits the process: nuclear excitation is read out via IC electrons from the isomer's rapid decay. The $12.3(3)\,\mu$s IC lifetime corresponds to a natural linewidth of $1/(2\pi\tau)\simeq13$~kHz and enables fast cycling; combined with the high nuclear density in a thin film, this yields an estimated short-term instability of about $2 \times 10^{-18}$ at 1~s~\cite{Elwell2025NatureThO2}. Rather than requiring VUV fluorescence from a high-bandgap transparent host, this approach turns the otherwise quenching IC channel into the readout signal and is compatible with low-bandgap or opaque materials such as ThO$_2$~\cite{Elwell2025NatureThO2}. In 2025, Elwell \textit{et al.} validated this path with laser-driven conversion-electron Mössbauer spectroscopy in $\text{ThO}_2$~\cite{Elwell2025NatureThO2}, directly measuring the rapid IC decay. The simple photocurrent readout further opens the door to clock miniaturization.

\section{Nuclear Structure of $^{229}$Th}\label{sec:theory}

The extremely low excitation energy of the $^{229}\text{Th}$ isomer arises from a near-degeneracy between the ground state ($5/2^+$) and the isomeric state ($3/2^+$) within the Nilsson orbitals $5/2^+[633]$ and $3/2^+[631]$. This exceptionally small energy scale ($\sim 8\text{ eV}$) requires a near-perfect cancellation between various $\text{MeV}$-scale contributions to the total energy difference, placing a stringent demand on the precision of theoretical descriptions. Given the unknown nature of nuclear forces at short distances and the complexities of the many-body problem, no current theoretical framework can predict this $\sim 8~\mathrm{eV}$ transition energy with the required accuracy. Existing theoretical approaches therefore need to be assessed not only by the transition energy, but also through complementary observables, such as electromagnetic moments and transition probabilities, that provide independent benchmarks for the calculated wave functions. 

The Nilsson model provides widely used configuration assignments for the ground and isomeric states in terms of deformed single-particle orbitals.  
Incorporating Coriolis mixing allows for the estimation of interband transition matrix elements and hence the isomeric lifetime, since without $K$ mixing the $\Delta K=1$ isomeric transition would vanish in the axially symmetric limit. In the actinide region around $^{229}$Th, octupole correlations are also important, arising from the $\Delta l = \Delta j = 3$ coupling between opposite-parity orbitals near the Fermi surface, including the proton $\pi(f_{7/2}\leftrightarrow i_{13/2})$ and neutron $\nu(g_{9/2}\leftrightarrow j_{15/2})$ pairs~\cite{Wang2026CPB,Butler1996RevModPhys}. These correlations modify the parity content of the Nilsson wave functions and affect the $M1$ matrix elements~\cite{Minkov2019Phys.Rev.Lett.Theoretical}. \red{The coherent quadrupole--octupole mode (CQOM), combined with the deformed shell model (DSM), integrates collective vibrations and rotations with single-particle motion and} has successfully captured the role of octupole deformation in the low-lying structure of $^{229}$Th and provided a good description of the isomeric magnetic dipole moment~\cite{Minkov2017Phys.Rev.Lett.Reduced,Minkov2019Phys.Rev.Lett.Theoretical}.  
Furthermore, the Projected Shell Model (PSM), a microscopic model that diagonalizes an effective nuclear Hamiltonian in a symmetry-projected multi-quasiparticle basis, and self-consistent nuclear density functional theory (DFT) calculations, from the mean-field level to beyond-mean-field with symmetry restoration and configuration mixing, provide independent constraints on electromagnetic moments and transition probabilities~\cite{Chen2025Phys.Lett.BPSM,Minkov2024Phys.Rev.CHFBCS,Restrepo2026DFT,zhou2025microscopic}.
 
In this section, we discuss these approaches ranging from the Nilsson framework and the Coriolis mixing mechanism, through the CQOM-DSM method, to representative microscopic many-body calculations, and conclude with a systematic comparison with experimental data. Within the discussion of the Nilsson framework, we also briefly note that the $5/2^+[633]$ and $3/2^+[631]$ orbitals satisfy the asymptotic quantum-number relation for deformed pseudospin partners, raising the possibility of a dynamical-symmetry perspective on their near-degeneracy.

\subsection{Nilsson Model and Coriolis Mixing}

The Nilsson model~\cite{Nilsson1955MatFysMedd,Nilsson1060_NPA131-1} provides the natural framework for understanding the low-lying states of $^{229}$Th. It describes single-particle motion of nucleons in a deformed, axially symmetric modified oscillator potential with $\bm{l}\cdot\bm{s}$ (spin-orbit) and $\bm{l}^2$ (orbit-orbit) correction terms:
\begin{equation}
\begin{aligned}
    H_{\rm sp} = {} & \hbar\omega_0 \left( H_0 - \frac{2}{3}\varepsilon_2 \rho^2 P_2(\cos\theta) \right) \\
    & - \kappa\hbar\omega_0 \left[ 2\bm{l}\cdot\bm{s} + \mu(\bm{l}^2 - \langle\bm{l}^2\rangle_N) \right],
    \label{eq:Nilsson}
\end{aligned}
\end{equation}
where $H_0$ is the isotropic harmonic oscillator Hamiltonian, $\varepsilon_2$ is the quadrupole deformation parameter, $\rho$ is the dimensionless radial coordinate, and $\kappa$ and $\mu$ are shell-dependent phenomenological Nilsson parameters, with $\kappa$ setting the spin-orbit strength and $\mu$ controlling the relative strength of the $l^2$ correction. For well-deformed nuclei, the eigenstates are approximately characterized by the asymptotic quantum numbers $\Omega^\pi[Nn_z\Lambda]$, where $\Omega$ is the projection of the total angular momentum on the symmetry axis, $N$ is the principal oscillator quantum number, $n_z$ the number of quanta along the symmetry axis, and $\Lambda$ the projection of the orbital angular momentum~\cite{Nilsson1955MatFysMedd}.

For $^{229}$Th ($Z=90$, $N=139$), the configurations of the ground state and the isomeric state are primarily determined  by the unpaired neutron outside the even-even $N=138$ core.  Kroger and Reich~\cite{Kroger1976NuclPhysA} first studied the level scheme of $^{229}$Th via $\alpha$-decay spectroscopy of $^{233}$U. They observed rotational bands built on the $K = 5/2$ and $K = 3/2$ bandheads, and from the band structure and $\alpha$-decay hindrance factors determined the ground-state spin-parity as $I^\pi = 5/2^+$ and the first excited state as $I^\pi = 3/2^+$, assigning the Nilsson orbitals as $5/2^+[633]$ and $3/2^+[631]$. These assignments have been consistently confirmed by subsequent $\gamma$-ray spectroscopy~\cite{Gulda2002NuclPhysA,Barci2003Phys.Rev.CNucleara,Ruchowska2006Phys.Rev.CNuclear}.

\begin{figure}[t]
\centering
\includegraphics[width=\columnwidth]{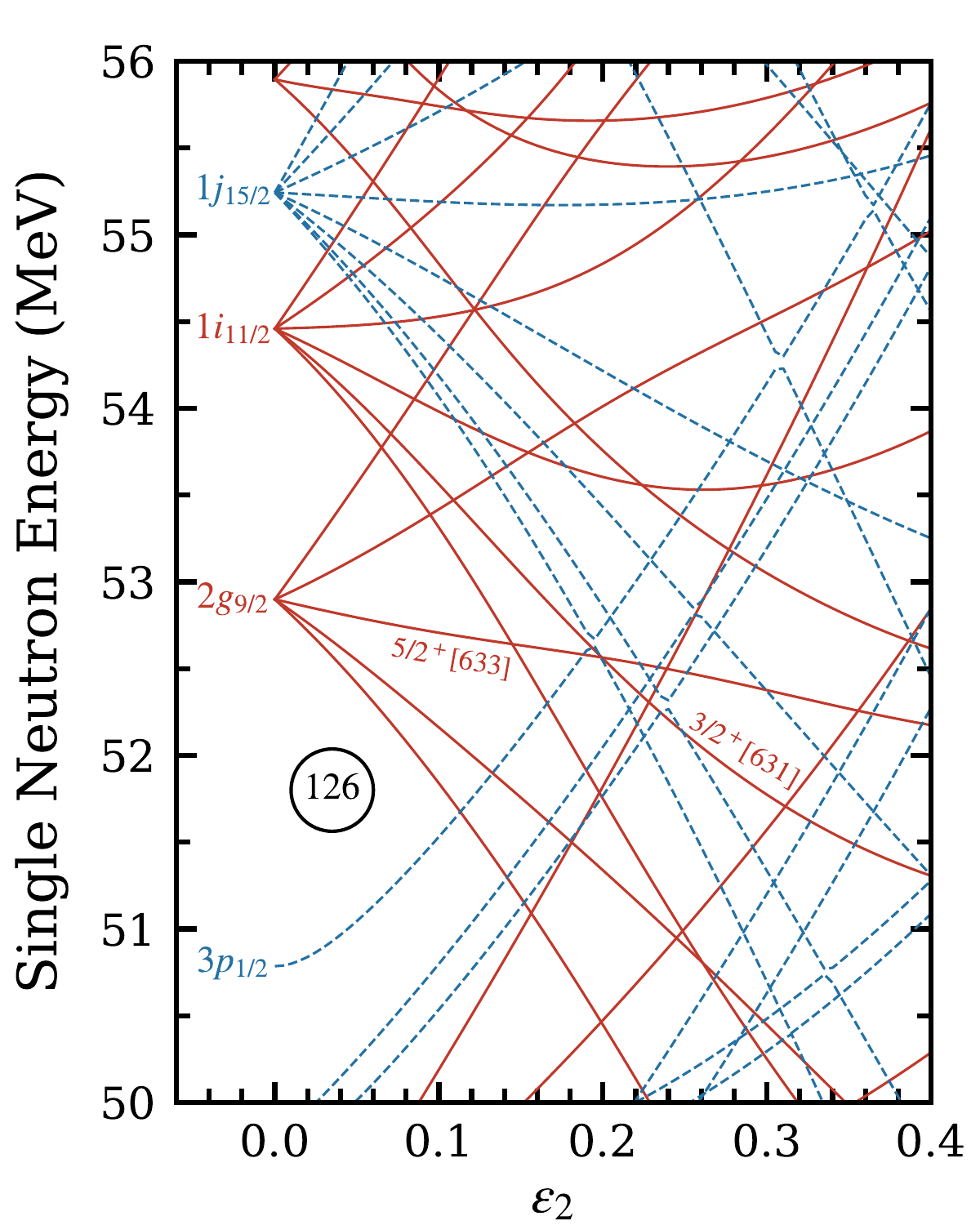}
\caption{Nilsson single neutron energy levels for $^{229}$Th as a function of quadrupole deformation $\varepsilon_2$, calculated with the reflection-symmetric modified oscillator Hamiltonian using the Bengtsson--Ragnarsson parameters~\cite{Bengtsson1985NuclPhysA}. Solid red (dashed blue) lines denote positive- (negative-) parity orbitals. The labels indicate the spherical orbitals at $\varepsilon_2=0$ and the spectroscopic Nilsson assignments $5/2[633]$ and $3/2[631]$. The circled number marks the $N=126$ shell gap.  }
\label{fig:nilsson}
\end{figure}

Experimental constraints on the ground-state deformation are provided by Coulomb excitation with $\alpha$ particles~\cite{Bemis1988Coulomb}, which yields an intrinsic quadrupole moment $Q_{20} = 8.816 \pm 0.090~e\,{\rm b}$ and an intrinsic hexadecapole moment $Q_{40} = 3.69 \pm 0.72~e\,{\rm b}^2$. For a uniformly charged axially deformed nucleus, the intrinsic electric quadrupole moment can be expressed in terms of the quadrupole deformation $\beta_2$ as~\cite{BohrMottelson1975} 
\begin{equation}
    Q_{20} = \frac{3}{\sqrt{5\pi}} Z R_0^2 \beta_2 (1 + 0.36\,\beta_2 + \cdots),
    \label{eq:Q20}
\end{equation}
with $R_0 = r_0 A^{1/3}$ and $r_0 = 1.2$~fm, one obtains $\beta_2 \approx 0.22$ (corresponding to $\varepsilon_2 \approx 0.21$) for $^{229}$Th. As shown in Fig.~\ref{fig:nilsson}, the spectroscopically assigned $5/2^+[633]$ and $3/2^+[631]$ orbitals approach each other with increasing deformation; in the present calculation with the Bengtsson--Ragnarsson parameters~\cite{Bengtsson1985NuclPhysA}, their crossing occurs at $\varepsilon_2 \approx 0.20$, close to the deformation inferred from the quadrupole moment. However, this schematic reflection-symmetric Nilsson calculation does not place the last neutron directly in these two orbitals. The diagram should therefore be viewed as illustrating the deformation-driven near-degeneracy of the assigned configurations, while a quantitative determination of the Fermi-level ordering in $^{229}$Th requires ingredients beyond this schematic reflection-symmetric Nilsson calculation, such as pairing, octupole deformation, configuration mixing, and possible adjustments of the Nilsson parameters. A related indication is provided by the PSM calculation of Ref.~\cite{Chen2025Phys.Lett.BPSM}, which uses slightly adjusted neutron $N=6$ Nilsson parameters and includes octupole deformation; after angular-momentum and parity projection together with configuration mixing, the $5/2[633]$ configuration becomes dominant in the ground state while the $3/2[631]$ configuration becomes dominant in the isomeric state.

The near-degeneracy of the assigned $3/2^+[631]$ and $5/2^+[633]$ Nilsson orbitals is reminiscent of the pseudospin doublets known in nuclear single-particle spectra. Pseudospin symmetry was introduced to describe the empirical quasi-degeneracy of normal-parity orbitals $(n,l,j=l+1/2)$ and $(n-1,l+2,j=l+3/2)$, which can be relabeled by a common pseudo-orbital angular momentum $\tilde l=l+1$ coupled to pseudospin $\tilde s=1/2$~\cite{Hecht1969NuclPhysAGeneralized,Arima1969PhysLettBPseudo,Liang2015_PR570-1}. \red{In axially deformed, reflection-symmetric mean fields,} the analogous Nilsson partners have asymptotic labels $[Nn_z\Lambda](\Omega=\Lambda+1/2)$ and $[Nn_z,\Lambda+2](\Omega=\Lambda+3/2)$, or equivalently a common pseudo-orbital projection $\tilde{\Lambda}=\Lambda+1$ with $\Omega=\tilde{\Lambda}\pm1/2$~\cite{Bohr1982_PS26-267,Ginocchio2004PRC69-034303}. The $3/2^+[631]$ and $5/2^+[633]$ assignments satisfy this deformed pseudospin-partner pattern, with $\tilde{\Lambda}=2$. This correspondence provides a useful way to formulate the structural origin problem of the $^{229}$Th isomer: the exceptionally small bandhead separation may be viewed in relation to approximate pseudospin symmetry and its breaking in an actinide mean field. Existing studies have established the pseudospin classification and its tests in \red{axially deformed, reflection-symmetric} mean fields~\cite{Bohr1982_PS26-267,Ginocchio2004PRC69-034303,Liang2015_PR570-1}, \red{while whether this interpretation remains quantitatively valid after octupole correlations and the associated parity mixing are included has yet to be clarified for $^{229}$Th.}

In the axially symmetric Nilsson model, the $3/2^+[631] \to 5/2^+[633]$ $M1$ transition is forbidden at leading order. Although the $K$-selection rule formally permits $M1$ transitions with $\Delta K = 1$, the leading-order intrinsic matrix element is suppressed because the two orbitals differ by $\Delta\Lambda = 2$ ($\Lambda = 1$ for [631] vs.\ $\Lambda = 3$ for [633]), while the single-particle $M1$ operator can connect states with $\Delta\Lambda \leq 1$ only~\cite{Dykhne1998JetpLett.Matrix}. However, the rotation of the nuclear core introduces a Coriolis interaction that mixes the $K = 3/2$ and $K = 5/2$ rotational bands~\cite{BohrMottelson1975}, introducing admixtures that lift the $\Delta\Lambda$ forbiddenness and generate a nonzero $M1$ matrix element.

The first quantitative estimate exploiting this mechanism was provided by Dykhne and Tkalya~\cite{Dykhne1998JetpLett.Matrix}. Working within the rotational model in first-order perturbation theory, they introduced the following phenomenological expression for the interband $M1$ matrix element:
\begin{equation}
\begin{aligned}
&\langle K_f I_f \| \mathcal{M}(M1)\| K_i I_i\rangle\\
= &\sqrt{2I_i+1}\,
\langle I_i K_i 1\,1 | I_f K_f\rangle
\left(\mathcal{A} + \mathcal{B}[I_f(I_f+1)-I_i(I_i+1)]\right),
\end{aligned}
\label{eq:M1matrix}
\end{equation}
where $\langle I_i K_i 1\,1 | I_f K_f\rangle$ is the Clebsch--Gordan coefficient;  $\mathcal{A}$ is extracted from the 25.3 keV interband $M1$ transition $9/2^+(K=5/2)\to 7/2^+(K=3/2)$ in $^{229}$Th~\cite{Kroger1976NuclPhysA}. $\mathcal{B}$ is determined from the experimentally inferred band parameters $g_K-g_R$ for the $K=5/2$ and $K=3/2$ bands, together with a dimensionless Coriolis mixing matrix element of 0.052 extracted from the interband $E2$ transition probability~\cite{Dykhne1998JetpLett.Matrix}. Here $g_K$ is the intrinsic gyromagnetic factor of the band, while $g_R$ denotes the collective rotational gyromagnetic factor of the core. The corresponding reduced transition probability is
\begin{equation}
B(M1; I_iK_i \to I_fK_f)
=
\frac{1}{2I_i+1}
\left|
\langle K_f I_f \| \mathcal{M}(M1)\| K_i I_i\rangle
\right|^2 .
\end{equation}
With these inputs, they obtained 
\begin{equation}
    B(M1; 3/2^+ \to 5/2^+) = 0.086\, \mu_N^2=0.048~\mathrm{W.u.},
\end{equation}
corresponding to a radiative lifetime of $\sim 2.5$ hours at the then-accepted transition energy of 3.5 eV. The predicted isomeric magnetic moment is $\mu_{\rm IS} \approx -0.076\,\mu_N$.

While this pioneering estimate provided the first quantitative picture, subsequent measurements revealed its limitations. The current experimental value of the isomeric magnetic moment, $\mu_{\rm IS}=-0.378(8)\,\mu_N$~\cite{Yamaguchi2024natureLaser}, is approximately five times larger in magnitude than the Dykhne--Tkalya prediction. A refined phenomenological analysis by Tkalya \textit{et al.}~\cite{Tkalya2015PRC}, using additional experimental $\gamma$-ray intensities from the $5/2^+[633]$ and $3/2^+[631]$ rotational bands, constrained the radiative lifetime to $0.66 \times 10^6\,{\rm s\,eV^3}/\omega^3 \leqslant \tau \leqslant 2.2 \times 10^6\,{\rm s\,eV^3}/\omega^3$ at the 95\% confidence level, but did not resolve the $\mu_{\rm IS}$ discrepancy. This indicates that Coriolis mixing alone, applied within a reflection-symmetric Nilsson framework, is insufficient to describe the electromagnetic properties of the isomeric state. These limitations motivate the inclusion of parity mixing induced by octupole deformation, as implemented in the CQOM-DSM framework discussed below.

\begin{figure*}[ht!]
\centering
\includegraphics[width=12 cm]{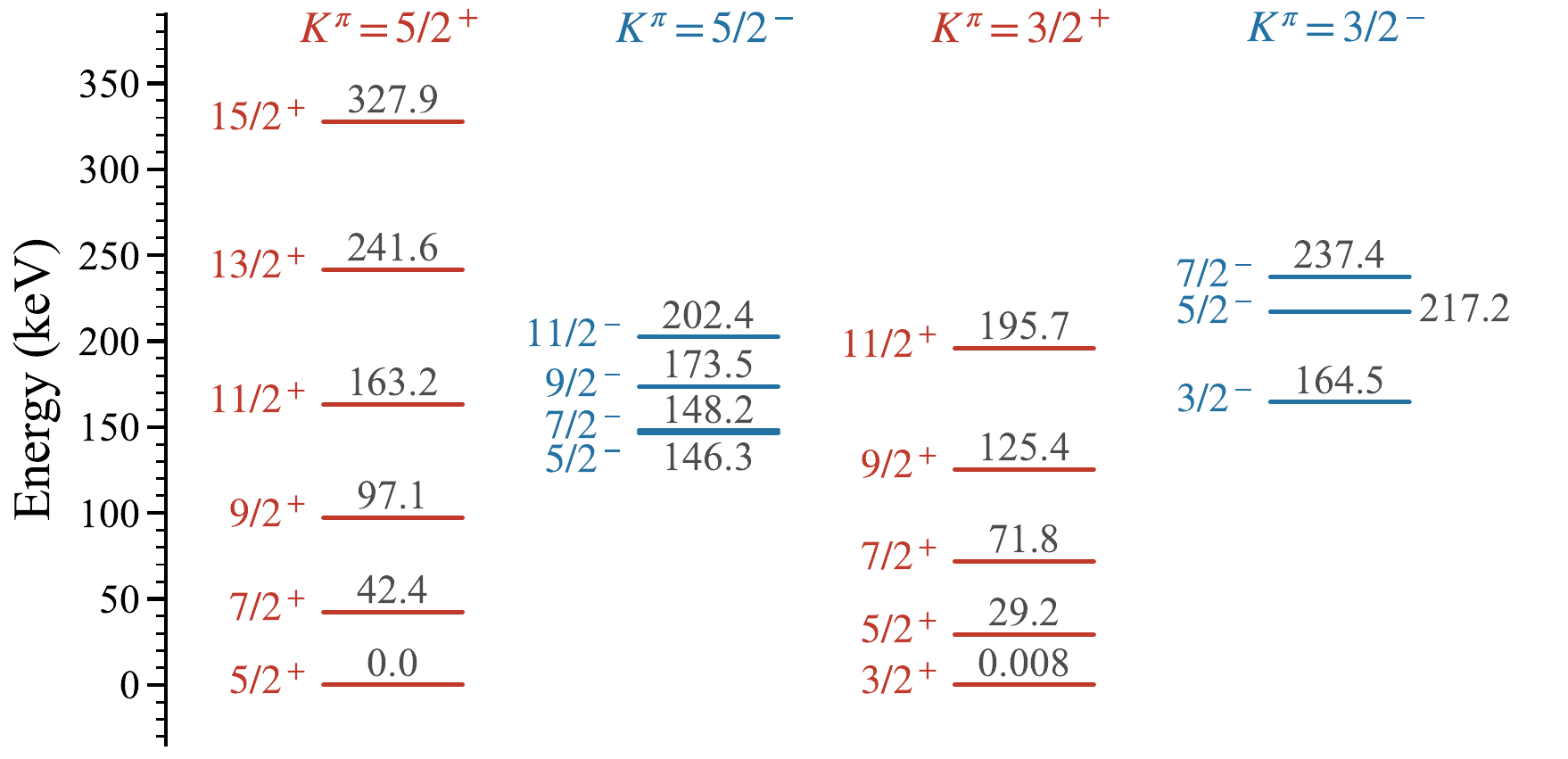}
\caption{Low-energy level scheme of $^{229}$Th showing the four rotational bands relevant to the isomeric transition: the $K^\pi = 5/2^+$ ground-state band and its $K^\pi = 5/2^-$ parity partner (left), and the $K^\pi = 3/2^+$ isomeric band and its $K^\pi = 3/2^-$ parity partner (right). Positive-parity bands are shown in red, negative-parity bands in blue. The isomeric $M1/E2$ transition ($\sim 8$ eV) connects the two bandheads. Level energies are from ENSDF~\cite{ENSDF}; band assignments follow Refs.~\cite{Kroger1976NuclPhysA,Ruchowska2006Phys.Rev.CNuclear}.}
\label{fig:level_scheme}
\end{figure*}

\subsection{\texorpdfstring{Coherent Quadrupole-Octupole \red{Mode} \red{Combined} with the Deformed Shell Model}{Coherent Quadrupole-Octupole Mode Combined with the Deformed Shell Model}}

In 2002, Gulda \textit{et al.}~\cite{Gulda2002NuclPhysA} identified two octupole-correlated parity partner bands in $^{229}$Th, $K^\pi=5/2^\pm$ and $K^\pi=3/2^\pm$, providing spectroscopic evidence for octupole correlations. Figure~\ref{fig:level_scheme} summarizes the relevant low-energy bands. 
In the presence of octupole deformation, the reflection-asymmetric potential mixes opposite-parity components into the intrinsic single-particle wave functions, significantly modifying the magnetic moment and transition matrix elements compared to the pure Nilsson-plus-Coriolis picture.

To incorporate octupole correlations, Minkov \textit{et al.}\ developed \red{a collective model of coherent quadrupole--octupole motion}~\cite{Minkov2006Phys.Rev.CCQOM,Minkov2007Phys.Rev.COddA} and later \red{combined it with the deformed shell model (DSM) to form the CQOM-DSM approach} for application to $^{229}$Th~\cite{Minkov2017Phys.Rev.Lett.Reduced,Minkov2019Phys.Rev.Lett.Theoretical}. The model Hamiltonian takes the form~\cite{Minkov2017Phys.Rev.Lett.Reduced}
\begin{equation}
    H = H_{\rm SP} + H_{\rm pair} + H_{\rm QO} + H_{\rm Coriol},
    \label{eq:CQOM-DSM}
\end{equation}
where $H_{\rm SP}$ is the single-particle Hamiltonian with a reflection-asymmetric Woods-Saxon potential including axial quadrupole, octupole, and higher-multipolarity deformations; $H_{\rm pair}$ is the BCS pairing Hamiltonian; $H_{\rm QO}$ describes the coherent quadrupole-octupole vibration-rotation motion of the even-even core; and $H_{\rm Coriol}$ is the Coriolis interaction between the unpaired nucleon and the core, treated as a perturbation~\cite{Minkov2017Phys.Rev.Lett.Reduced}. This framework naturally produces quasiparity doublet spectra, consistent with the experimentally observed $K^\pi = 5/2^\pm$ and $3/2^\pm$ parity partner bands in $^{229}$Th~\cite{Ruchowska2006Phys.Rev.CNuclear}.

In Ref.~\cite{Minkov2017Phys.Rev.Lett.Reduced}, the quadrupole and octupole deformations in the DSM are determined by requiring that the $5/2[633]$ and $3/2[631]$ orbitals appear as the last occupied and first unoccupied neutron states with the smallest possible spacing. With $\beta_2$ scanned within the experimental window set by the neighboring even-even nuclei $^{228}$Th ($\beta_2 = 0.230$) and $^{230}$Th ($\beta_2 = 0.244$)~\cite{Raman2001ADNDT} and $\beta_3$ varied freely, this condition is met at $\beta_2 = 0.240$ and $\beta_3 = 0.115$; the correct orbital ordering is obtained only at nonzero $\beta_3$. With these deformations and the CQOM parameters fitted to the low-lying spectrum, the model reproduces the positive- and negative-parity level scheme of $^{229}$Th below 400 keV and predicts the isomeric $M1$ transition probability as
\begin{equation}
    B(M1; 3/2^+ \to 5/2^+) = 0.006\text{--}0.008 \text{ W.u.},
\end{equation}
nearly an order of magnitude smaller than the Dykhne--Tkalya estimate of 0.048 W.u.\ and about half the quasiparticle-plus-phonon model value, $B(M1)=0.025\,\mu_N^2 \simeq 0.014$ W.u.~\cite{Ruchowska2006Phys.Rev.CNuclear}. This reduction was proposed as a possible explanation for the lack of experimental success at that time in driving or observing the radiative channel~\cite{Minkov2017Phys.Rev.Lett.Reduced}. The model also predicted $B(E2; 3/2^+ \to 5/2^+) \approx 20$--$30$ W.u., suggesting that the $E2$ channel may make a non-negligible contribution to internal conversion in certain electronic configurations. 

A more stringent test came in 2019~\cite{Minkov2019Phys.Rev.Lett.Theoretical}, motivated by the 2018 laser-spectroscopy determination of the isomeric magnetic moment, $\mu_{\rm IS}=-0.37(6)\,\mu_N$~\cite{Thielking2018NatureLasera}. Within the same model framework, the magnetic moments are evaluated for several values of the quenching factor $q_R$ of the collective gyromagnetic ratio $g_R = q_R Z/A$~\cite{Minkov2019Phys.Rev.Lett.Theoretical}:
\begin{itemize}
  \item $q_R = 0.6$: $\mu_{\rm IS} = -0.347\,\mu_N$, $\mu_{\rm GS} = 0.528\,\mu_N$
  \item $q_R = 0.7$: $\mu_{\rm IS} = -0.323\,\mu_N$, $\mu_{\rm GS} = 0.559\,\mu_N$
  \item $q_R = 0.8$: $\mu_{\rm IS} = -0.300\,\mu_N$, $\mu_{\rm GS} = 0.591\,\mu_N$
\end{itemize}
For $q_R = 0.6$--$0.7$, the calculated isomeric magnetic moment agrees with the measured value, whereas the calculated ground-state magnetic moment $\mu_{\rm GS}$ overestimates the experimental value of $0.36\,\mu_N$~\cite{Minkov2019Phys.Rev.Lett.Theoretical} by a factor of 1.5--1.6. The analysis further showed that Coriolis mixing is crucial for the isomeric transition rates, but has only a small effect on $\mu_{\rm GS}$ and a negligible effect on $\mu_{\rm IS}$; the quenching of $g_R$ instead controls the agreement of the calculated magnetic moments with experiment~\cite{Minkov2019Phys.Rev.Lett.Theoretical}. For $q_R=0.6$, the isomeric $B(M1)$ value is reduced to $0.0056$ W.u. 

With $B(M1) \approx 0.005$--$0.008$ W.u.\ and the now-established transition energy of $\approx 8$ eV, the predicted $M1$ radiative half-life is roughly $(5$--$8)\times 10^3$ s, about a factor of $3$--$4$ longer than the vacuum half-life $T_{1/2}=1740(50)$ s reported by Tiedau \textit{et al.}~\cite{Tiedau2024physrevlettLaser}.

\subsection{Projected Shell Model}

The Projected Shell Model (PSM)~\cite{Hara1995IJMPE,Sun1996PhysRep} starts from a deformed Nilsson mean field with BCS pairing, constructs multi-quasiparticle (qp) configurations, and restores rotational symmetry by angular-momentum projection before diagonalizing the nuclear Hamiltonian. In the $^{229}$Th application, the inclusion of octupole deformation additionally requires parity projection~\cite{Chen2025Phys.Lett.BPSM}. The Hamiltonian diagonalized in the projected basis is taken as
\begin{equation}
\begin{aligned}
  \hat H ={}& \hat H_0
  - \frac{1}{2}\sum_{L=2}^{4}\chi_L
  \sum_{\omega=-L}^{L}\hat Q^{\dagger}_{L\omega}\hat Q_{L\omega}- G_M\,\hat P^{\dagger}\hat P \\
  &
  - G_Q\sum_{\omega=-2}^{2}\hat P^{\dagger}_{2\omega}\hat P_{2\omega}.
\end{aligned}
\label{eq:PSM-H}
\end{equation}
where $\hat H_0$ is a deformed Nilsson single-particle Hamiltonian of the form given in Eq.~\eqref{eq:Nilsson}, extended to include octupole ($\varepsilon_3$) and hexadecapole ($\varepsilon_4$) deformations in addition to $\varepsilon_2$. The second term represents separable multipole-multipole interactions in the particle-hole channel, where $\hat Q_{L\omega}$ is the multipole operator and $L=2$, 3, and 4 correspond to the quadrupole-quadrupole, octupole-octupole, and hexadecapole-hexadecapole interactions, respectively. The strengths $\chi_L$ are fixed self-consistently so that the multipole-multipole interactions reproduce the deformations prescribed in $\hat H_0$, rather than being fitted independently. The last two terms describe monopole and quadrupole pairing interactions in the particle-particle channel: $\hat P^\dagger$ creates a monopole pair, while $\hat P_{2\omega}^\dagger$ creates a quadrupole pair. The monopole-pairing strength is parametrized as $G_M=[G_1\mp G_2(N_n-Z)/A]/A$,  with the plus (minus) sign for protons (neutrons), and the values $G_1=16.32~\mathrm{MeV}$ and $G_2=11.81~\mathrm{MeV}$ are slightly modified from those optimized for neighboring even-even thorium isotopes. The quadrupole-pairing strength is taken as $G_Q/G_M=0.13$ for this mass region~\cite{Chen2025Phys.Lett.BPSM}. With these ingredients, symmetry restoration and configuration mixing yield correlated laboratory-frame wave functions for calculating spectroscopic observables.

Chen \textit{et al.}~\cite{Chen2025Phys.Lett.BPSM} applied the PSM to $^{229}$Th, incorporating for the first time octupole deformation ($\varepsilon_3$) alongside quadrupole and hexadecapole degrees of freedom with parity projection in this framework. The model space includes three major shells for both neutrons ($N = 5, 6, 7$) and protons ($N = 4, 5, 6$), with configurations up to 3-qp (5-qp verified to have negligible effect).   The Nilsson deformation parameters $\varepsilon_2 = 0.19$, $\varepsilon_3 = 0.066$, and $\varepsilon_4 = -0.089$ follow the deformation systematics adopted in Ref.~\cite{Chen2025Phys.Lett.BPSM}, with the neutron $N=6$ Nilsson parameters ($\kappa$, $\mu$) slightly adjusted to reproduce the band-head energies.

The calculation reproduces the low-energy level scheme of $^{229}$Th, including the ground-state band, the isomer band, and two low-lying negative-parity bands. Analysis of the wave functions reveals that the ground state is predominantly (77.4\%) the $5/2[633]$ configuration, while the isomeric state is 82.8\% the $3/2[631]$ configuration~\cite{Chen2025Phys.Lett.BPSM}. Notably, at the mean-field level, the neutron Fermi surface lies at the $3/2[631]$ one-quasiparticle state, while the $5/2[633]$ state is higher by 75 keV. After angular-momentum and parity projection followed by configuration mixing, the $5/2[633]$ configuration becomes the dominant component of the ground state and the $3/2[631]$ configuration becomes that of the low-lying isomer, with the quadrupole-pairing term playing an important role in this rearrangement~\cite{Chen2025Phys.Lett.BPSM}.

For the isomeric $M1$ transition, using the bare neutron spin $g$-factor and the proton spin $g$-factor quenched by 0.85, the model yields
\begin{equation}
    B(M1; 3/2^+ \to 5/2^+) = 0.024 \text{ W.u.},
\end{equation} 
consistent with the values inferred from recent radiative-lifetime measurements, $B(M1)\sim 0.017$--$0.030$ W.u., assuming a purely radiative $M1$ decay and applying the refractive-index correction in crystals~\cite{Kraemer2023natureObservation,Tiedau2024physrevlettLaser,Elwell2024physrevlettLaser}. The PSM predicts $B(E2; 3/2^+ \to 5/2^+) = 8.74$ or $10.95$ W.u., depending on the adopted effective charges, smaller than the CQOM-DSM range of $20$--$30$ W.u., suggesting that the $E2$ channel may be less important for the isomeric decay than previously estimated~\cite{Minkov2017Phys.Rev.Lett.Reduced,Chen2025Phys.Lett.BPSM}. The analysis of the reduced one-body transition density shows that the orbital contributions entering the $M1$ matrix element are highly dispersed, with sizable components not only near but also far from the Fermi surface. This indicates that a large model space is indispensable for reliable  calculations of the $^{229\mathrm{m}}\mathrm{Th}$ decay~\cite{Chen2025Phys.Lett.BPSM}. 

\subsection{Skyrme Hartree-Fock-BCS}
The Skyrme Hartree-Fock-BCS (HFBCS) approach, based on the self-consistent Skyrme energy-density-functional framework~\cite{Bender2003RevModPhysSelfConsistent,Stone2007ProgPartNuclPhysSkyrme}, incorporates the polarization of the even-even core by the unpaired nucleon through self-consistent blocking and the associated time-odd mean-field terms in odd-mass nuclei~\cite{Pototzky2010EurPhysJATimeOdd,Minkov2024Phys.Rev.CHFBCS}. By allowing reflection-asymmetric mean-field solutions, it can include an axial octupole degree of freedom, which is particularly relevant for $^{229}$Th in light of nuclear-structure evidence for octupole correlations, including parity-partner bands and enhanced $E1$ transitions~\cite{Ruchowska2006Phys.Rev.CNuclear}. 
 
Minkov \textit{et al.}~\cite{Minkov2024Phys.Rev.CHFBCS} applied this framework to $^{229}$Th and neighboring nuclei using the SIII Skyrme parameterization~\cite{Beiner1975NuclPhysA}. The quadrupole and octupole deformations are not input parameters but emerge from the self-consistent variational procedure, yielding $\beta_2 \approx 0.213$--$0.216$ and $\beta_3 \approx 0.062$--$0.110$ for $^{229}$Th, depending on the blocked quasiparticle configuration~\cite{Minkov2024Phys.Rev.CHFBCS}. A central finding is that the near-degeneracy of the $5/2[633]$ and $3/2[631]$ neutron orbitals is reproduced only when octupole deformation is included in the mean field: without $\beta_3 \neq 0$, the orbital ordering is incorrect or the splitting is too large~\cite{Minkov2024Phys.Rev.CHFBCS}. This provides independent confirmation, within a self-consistent framework, of the essential role of octupole correlations in this nucleus. The predicted excitation energy of the isomeric state remains in the keV range rather than the eV range, as expected since current self-consistent mean-field theories cannot reproduce an eV-scale energy splitting with an accuracy of order $10^{-9}$ relative to the total nuclear binding energy ($\sim$1748 MeV~\cite{Wang2021ChinesePhys.CAME}). 

For electromagnetic moments, the calculation yields $\mu_{\rm GS}\approx 0.42$--$0.43\,\mu_N$, closer to the hyperfine-derived value $0.366(6)\,\mu_N$ than the reflection-symmetric result, while $\mu_{\rm IS}\approx -0.27\,\mu_N$ underestimates the measured $-0.378(8)\,\mu_N$~\cite{Porsev2021PhysRevLettHyperfine,Yamaguchi2024natureLaser}.  For the ground-state configuration, the HFBCS calculation gives $g_R=0.281$--$0.303$, corresponding to an effective attenuation relative to $Z/A \approx 0.393$ that is obtained self-consistently rather than introduced as a phenomenological parameter. This contrasts with the CQOM-DSM treatment, where $g_R=q_R Z/A$ with $q_R=0.6$--$0.8$ is imposed phenomenologically. The lower $\mu_{\rm GS}$ in HFBCS compared with the CQOM-DSM values may reflect the self-consistent treatment of core polarization~\cite{Minkov2024Phys.Rev.CHFBCS}.

A limitation of the present HFBCS approach is that it does not yet include Coriolis coupling between the odd neutron and the core, and hence lacks the $K$-mixing needed to calculate the $B(M1)$ transition probability between the isomeric and ground states~\cite{Minkov2024Phys.Rev.CHFBCS}. A particle-core extension built on the self-consistent solutions would be required for transition rates.

\subsection{Multi-Reference Nuclear DFT}

Building on the self-consistent Skyrme mean-field framework as the HFBCS calculation discussed above, but extending it to a multi-reference (configuration-interaction) treatment with parity and angular-momentum projection, Restrepo-Giraldo \textit{et al.}~\cite{Restrepo2026DFT} presented the first  multi-reference nuclear density functional theory (MR-DFT) calculation of the electromagnetic properties of $^{229}$Th. The calculation employs a multi-step procedure:  (1) self-consistent determination of octupole-deformed mean-field solutions including pairing correlations for the three lowest blocked quasiparticle configurations for each of $\Omega$ value in $^{229}$Th, providing a $6 \times 6$ basis for the subsequent configuration mixing; (2) parity and angular-momentum projection of each configuration; (3) configuration mixing via the Hill-Wheeler equation; and (4) evaluation of $M1$ and $E2$ matrix elements between the mixed $5/2^+$ and $3/2^+$ states. Seven Skyrme functionals (SkX$_c$, SkM*, UNEDF0, UNEDF1, SIII, SkO', SLy4) are employed to assess systematic uncertainties.  Since these functionals were not adjusted to octupole degrees of freedom, a linear regression anchored to the measured intrinsic octupole moments of $^{226}$Ra and $^{230}$Th is used to extrapolate the predictions~\cite{Restrepo2026DFT}.

A key methodological finding concerns the time-odd terms of the mean field, which are solely responsible for angular-momentum polarization of the core. Without these terms, $B(M1)$ is overestimated by an order of magnitude; with them included, the regression yields~\cite{Restrepo2026DFT}
\begin{equation}
  B(M1; 3/2^+ \to 5/2^+) = 0.04(3)\,\mu_N^2
  \simeq 0.022(17)\,\text{W.u.},
\end{equation}
consistent with the experimental value of $0.022$ W.u.~\cite{Tiedau2024physrevlettLaser}. The quoted uncertainty reflects the spread across the Skyrme functionals propagated through the regression. 
For the $^{226}$Ra-calibrated regression, the predicted spectroscopic electric quadrupole moments are $Q_{\rm GS}^{s}=3.01(7)\,e\,\mathrm{b}$ and $Q_{\rm IS}^{s}=1.69(6)\,e\,\mathrm{b}$~\cite{Restrepo2026DFT}, in good agreement with the ground-state value $Q_{\rm GS}^{s}=3.11(2)\,e\, \mathrm{b}$ deduced from hyperfine-structure measurements and atomic-structure calculations~\cite{Porsev2021PhysRevLettHyperfine}, and with the isomeric value $Q_{\rm IS}^{s}=1.77(2)\,e\,\mathrm{b}$ inferred from laser spectroscopy~\cite{Yamaguchi2024natureLaser}. 
However, the isomeric magnetic moment $\mu_{\rm IS} = -0.1(1)\,\mu_N$ underestimates the experimental $-0.378(8)\,\mu_N$ by a factor of $\sim 3$, while $\mu_{\rm GS} \approx 0.4(3)\,\mu_N$ is consistent with the experimental value within the large
  theoretical uncertainty~\cite{Restrepo2026DFT}.

The study also shows that configuration mixing has a limited effect on $B(M1)$: within the octupole-deformed calculations, the results obtained from the individual $5/2[633]$ and $3/2[631]$ blocked configurations are nearly identical to those from the mixed calculation. In contrast, parity-conserving calculations that neglect octupole deformation overestimate $B(M1)$ and disagree significantly with experiment, reinforcing the conclusion that reflection-asymmetric correlations are fundamental to the isomeric transition~\cite{Restrepo2026DFT}. The large spread of predictions across Skyrme functionals points to the need for a systematic adjustment of octupole degrees of freedom in future functional parametrizations.

\subsection{Multi-Reference Covariant DFT}

Covariant density functional theory (CDFT), also commonly referred to as relativistic density functional theory (RDFT), self-consistently includes relativistic effects and provides a microscopic framework for nuclear many-body systems~\cite{Serot1986_ANP16-1,Reinhard1989_RPP52-439,Ring1996_PPNP37-193,Vretenar2005_PR409-101,Meng2006_PPNP57-470,Paar2007_RPP70-691,Niksic2011_PPNP66-519,Meng_2015}. In CDFT/RDFT, nuclear saturation and the spin-orbit structure arise naturally from large scalar and vector mean fields in a Lorentz-covariant formulation. The effective interaction can be formulated either in a finite-range meson-exchange representation or in a zero-range point-coupling (PC) representation, with medium effects commonly incorporated through nonlinear terms or density-dependent coupling strengths~\cite{Vretenar2005_PR409-101,Niksic2011_PPNP66-519,Meng_2015}. This relativistic density-functional framework underlies both the multi-reference CDFT calculation discussed below and the RDFT input to the particle-rotor calculation discussed in the next subsection.

Zhou and Yao~\cite{zhou2025microscopic} applied multi-reference covariant density functional theory (MR-CDFT) to the low-lying states of several odd-mass nuclei relevant to atomic electric dipole moment (EDM) searches, including $^{229}$Th. The framework starts from self-consistent one-quasiparticle configurations generated with the  false quasiparticle vacuum scheme, employs the PC-PK1 relativistic functional~\cite{Zhao2010PRC_PCPK1}, and restores particle number, parity, and angular momentum before solving the Hill-Wheeler-Griffin equation with generator-coordinate mixing in the axial quadrupole-octupole deformation space $(\beta_2,\beta_3)$. In contrast to phenomenological particle-rotor descriptions, the electromagnetic transition operators are evaluated in the full single-particle space with bare proton and neutron charges.

For $^{229}$Th, the mean-field energy minimum is found at $(\beta_2,\beta_3)=(0.22,0.14)$, consistent with a strong quadrupole-octupole-deformed shape. After symmetry restoration, the projected $J^\pi=5/2^+$ minimum associated mainly with the neutron $5/2[633]$ configuration lies around $(\beta_2,\beta_3)=(0.22,0.15)$, whereas the $J^\pi=3/2^+$ isomeric configuration has a smaller octupole deformation and is predicted at an excitation energy of about 23 keV. As in other microscopic approaches, the eV-scale experimental splitting is not reproduced, but the calculation captures the main rotational-band structure: the ground-state band is predominantly a $K=5/2$ band with admixtures of the neutron $5/2[633]$, $5/2[503]$, and $5/2[752]$ components, while the isomeric band is dominated by the neutron $3/2[631]$ configuration.

The same calculation gives $\mu_{\rm GS}=0.32\,\mu_N$ and $Q_{\rm GS}^{s}=299\,e\,{\rm fm}^2=2.99\,e\,{\rm b}$ for the ground state, and $\mu_{\rm IS} =-0.48\,\mu_N$ and $Q_{\rm IS}^{s}=171\,e\,{\rm fm}^2=1.71\,e\,{\rm b}$ for the isomeric state, in reasonable agreement with available magnetic and quadrupole moment data~\cite{zhou2025microscopic}. For the isomeric transition, the calculation gives $B(E2;3/2_1^+\rightarrow 5/2_1^+)\simeq 5.6$ W.u. from the interband transition systematics shown in Fig.~21 of Ref.~\cite{zhou2025microscopic}, while the predicted $B(M1;3/2_1^+\rightarrow 5/2_1^+)=0.0045$ W.u. is several times smaller than the values inferred from recent radiative-lifetime measurements. Zhou and Yao also predict an additional, as-yet unobserved low-lying $5/2^-$ parity partner of the ground-state band at about 15 keV; if confirmed, this state would provide a substantially smaller opposite-parity energy denominator than the known $5/2^-$ bandhead at 146 keV and could enhance the Schiff moment.
 
\subsection{Relativistic DFT Combined with a Particle-Rotor Model}

Wang and Zhao~\cite{Wang2026CPB} employed RDFT solved on a three-dimensional lattice~\cite{Tanimura2015PTEP,Ren2017PRC_3Dlattice,Ren2019SciChina,Li2020PRC_3Dlattice,Xu2024PRC_3Dlattice}, using the PC-PK1 functional~\cite{Zhao2010PRC_PCPK1}. The resulting intrinsic configurations were then coupled to a reflection-asymmetric triaxial particle-rotor model (RAT-PRM)~\cite{Wang2019PLB_RATPRM} to describe the low-lying rotational structure of $^{229}$Th. 

The constrained RDFT calculation of Wang and Zhao yields a potential energy surface (PES) in the $(\beta_{20}, \beta_{30})$ plane with a global minimum at $\beta_{20} = 0.23$ and $\beta_{30} = 0.15$; the axial hexadecapole deformation $\beta_{40} = 0.18$ is obtained self-consistently. This result indicates a static axial quadrupole-octupole deformed ground state~\cite{Wang2026CPB}. Their analysis of the single-particle level evolution with $\beta_{30}$ attributes the microscopic origin of octupole correlations in $^{229}$Th to the neutron $g_{9/2}\leftrightarrow j_{15/2}$ and proton $f_{7/2}\leftrightarrow i_{13/2}$ octupole-driving couplings~\cite{Wang2026CPB}.
 
\begin{figure}[t]
\centering
\includegraphics[width=\columnwidth]{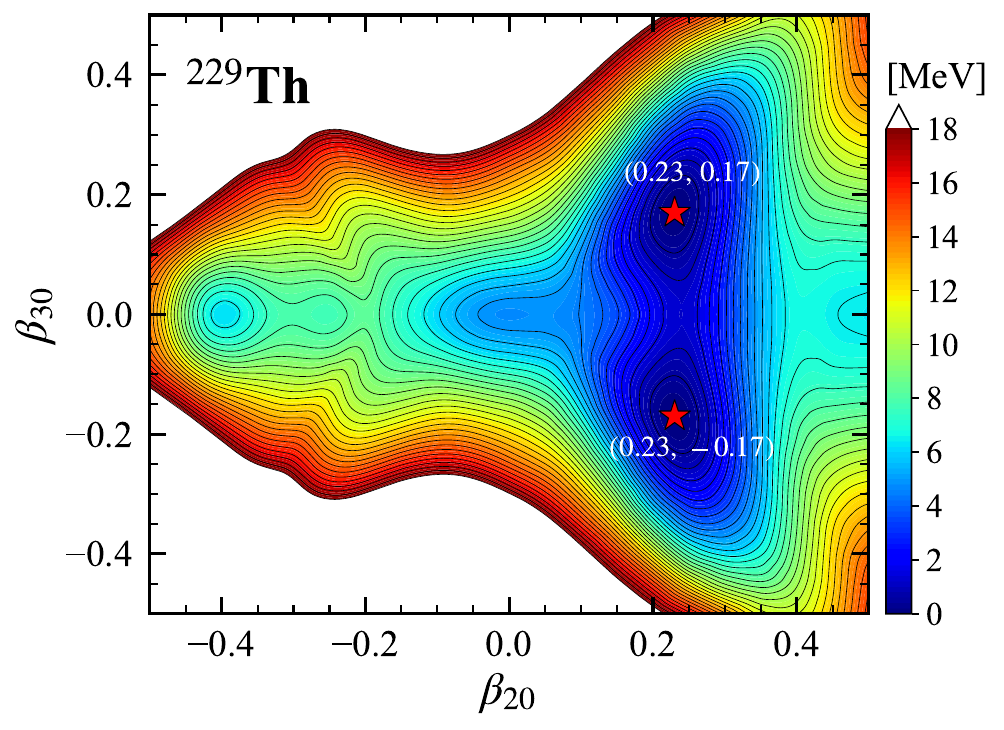}
\caption{Potential energy surface of $^{229}$Th in the $(\beta_{20},\,\beta_{30})$ plane, calculated in the present work with the MDC-RMF method~\cite{Lu2014PRC_MDCRMF,Zhou2016PhysScr_MDC} using the PC-PK1 functional~\cite{Zhao2010PRC_PCPK1}. The energy is normalized to the global minimum (red star), located at $(\beta_{20},\,\beta_{30}) = (0.23,\,0.17)$. The contour interval is 0.5 MeV; the white region indicates energies above 18 MeV.} 
\label{fig:PES}
\end{figure}

As a consistency check, we performed a multidimensionally-constrained relativistic mean-field (MDC-RMF) calculation~\cite{Lu2014PRC_MDCRMF,Zhou2016PhysScr_MDC} with the same PC-PK1 functional. The resulting PES, shown in Fig.~\ref{fig:PES}, has a global minimum at $(\beta_{20},\,\beta_{30}) = (0.23,\,0.17)$ with $\beta_{40} = 0.17$ obtained self-consistently. This close agreement with the 3D-lattice RDFT result supports the robustness of the predicted static octupole deformation in $^{229}$Th; the small difference in $\beta_{30}$ may reflect differences in numerical implementation, such as harmonic-oscillator basis expansion versus 3D lattice discretization.

Using the RDFT deformation as input, the RAT-PRM describes the rotational spectrum of the $K^\pi = 5/2^+$ ground-state band. The valence neutron space is spanned by the $1j_{15/2}$, $2g_{9/2}$, and $1i_{11/2}$ orbitals, and the pairing gap is taken in the empirical form $\Delta = 12/\sqrt{A}$ MeV. For the collective core, Wang and Zhao set the parity-splitting parameter to $E(0^-)=0$ and adopt a moment of inertia $\mathcal{J}_0 = 74\,\hbar^2\,{\rm MeV}^{-1}$ to reproduce the experimental energy spectrum. An ad hoc Coriolis attenuation factor $\xi = 0.4$ is introduced to reproduce the observed signature splitting. For electromagnetic transitions, the collective gyromagnetic ratio $g_R = Z/A\approx 0.393$ is adopted, while the valence-neutron gyromagnetic factor $g_n = -0.062$ is adjusted to reproduce the experimental ground-state magnetic moment~\cite{Wang2026CPB}.

The RAT-PRM calculation reproduces the available intraband $B(E2)$ values of the $K^\pi=5/2^+$ ground-state band reasonably well, with the $\Delta I = 1\hbar$ transitions slightly underestimated and the $\Delta I = 2\hbar$ transitions slightly overestimated~\cite{Wang2026CPB}. The corresponding intraband $B(M1)$ values are less well reproduced, differing from experiment by nearly an order of magnitude, consistent with the stronger sensitivity of $M1$ transitions to single-particle structure than of collective $E2$ transitions~\cite{Wang2026CPB}. The framework has not yet been extended to the $K^\pi = 3/2^+$ isomeric band or to the interband transition; such an extension  could provide a useful RDFT-constrained benchmark for the isomeric $B(M1)$ and $\mu_{\rm IS}$.

\subsection{Comparison with Experiment}

\red{Before comparing the calculated observables, Table~\ref{tab:theory-methods} summarizes the principal physical assumptions and treatments adopted in the theoretical approaches discussed above. The comparison highlights differences in whether and how octupole deformation is included, how rotational dynamics and symmetry restoration are incorporated, and how collective and single-particle $g$ factors and the $M1$ operator are treated. These methodological differences provide context for interpreting the theoretical predictions and their comparison with experiment below.}

\begin{table*}[ht!]
\centering
\caption{Comparison of the principal physical ingredients and treatments adopted in representative theoretical approaches to the low-lying structure of $^{229}$Th.}
\label{tab:theory-methods}
\scriptsize
\setlength{\tabcolsep}{1.8pt}
\renewcommand{\arraystretch}{1.25}

\newcommand{\modelcell}[2]{%
  \parbox[t]{#1}{%
    \raggedright #2\par
    \vspace{2pt}%
  }%
}

\newcommand{\markcell}[3]{%
  \parbox[t]{#1}{%
    \raggedright
    \hangindent=1.35em
    \hangafter=1
    \makebox[1.15em][c]{#2}\hspace{0.15em}#3\par
    \vspace{2pt}%
  }%
}

\begin{tabular*}{\textwidth}{@{\extracolsep{\fill}}lllllll@{}}
\hline\hline

\modelcell{2.05cm}{}
&
\modelcell{2.05cm}{\textbf{Model character}}
&
\modelcell{2.05cm}{\textbf{Octupole correlations}}
&
\modelcell{2.05cm}{\textbf{Particle number and parity}}
&
\modelcell{2.25cm}{\textbf{Rotation}}
&
\modelcell{2.65cm}{\textbf{Pairing and other correlations}}
&
\modelcell{2.65cm}{\textbf{Magnetic description}}
\\ \hline

\modelcell{2.05cm}{Dykhne \& Tkalya 1998~\cite{Dykhne1998JetpLett.Matrix}}
&
\modelcell{2.05cm}{Phenomenological rotational-model estimate}
&
\modelcell{2.05cm}{Not included}
&
\modelcell{2.05cm}{Reflection symmetric}
&
\modelcell{2.25cm}{First-order Coriolis mixing}
&
\modelcell{2.65cm}{No additional correlations}
&
\modelcell{2.65cm}{Empirical $g_R$ and $g_K-g_R$}
\\ \hline

\modelcell{2.05cm}{CQOM-DSM~\cite{Minkov2017Phys.Rev.Lett.Reduced,Minkov2019Phys.Rev.Lett.Theoretical}}
&
\modelcell{2.65cm}{Phenomenological QO collective + deformed single-particle model}
&
\modelcell{2.05cm}{Input $\beta_3$}
&
\modelcell{2.05cm}{PP}
&
\modelcell{2.25cm}{Perturbative Coriolis mixing}
&
\modelcell{2.65cm}{BCS pairing + collective QO motion}
&
\modelcell{2.65cm}{Phenomenologically quenched $g_R=q_R Z/A$}
\\ \hline

\modelcell{2.05cm}{PSM~\cite{Chen2025Phys.Lett.BPSM}}
&
\modelcell{2.05cm}{Effective shell-model Hamiltonian}
&
\modelcell{2.05cm}{Input $\varepsilon_3$}
&
\modelcell{2.05cm}{PP}
&
\modelcell{2.25cm}{AMP}
&
\modelcell{2.65cm}{BCS pairing + multi-qp configuration mixing}
&
\modelcell{2.65cm}{No separate $g_R$; fitted spin $g$ factors}
\\ \hline

\modelcell{2.05cm}{HFBCS~\cite{Minkov2024Phys.Rev.CHFBCS}}
&
\modelcell{2.05cm}{Self-consistent Skyrme}
&
\modelcell{2.05cm}{Self-consistent}
&
\modelcell{2.05cm}{Intrinsic parity mixing; no PP}
&
\modelcell{2.25cm}{No Coriolis mixing or AMP}
&
\modelcell{2.65cm}{BCS pairing + time-odd core polarization}
&
\modelcell{2.65cm}{Microscopic $g_R$ from cranking calculation}
\\ \hline

\modelcell{2.05cm}{MR-DFT~\cite{Restrepo2026DFT}}
&
\modelcell{2.05cm}{Self-consistent Skyrme}
&
\modelcell{2.05cm}{Self-consistent}
&
\modelcell{2.05cm}{PP}
&
\modelcell{2.05cm}{AMP}
&
\modelcell{2.65cm}{Volume pairing + configuration mixing + time-odd polarization}
&
\modelcell{2.65cm}{No separate $g_R$; microscopic $M1$ operator}
\\ \hline

\modelcell{2.05cm}{MR-CDFT~\cite{zhou2025microscopic}}
&
\modelcell{2.05cm}{Self-consistent covariant}
&
\modelcell{2.05cm}{Self-consistent}
&
\modelcell{2.05cm}{PNP + PP}
&
\modelcell{2.05cm}{AMP}
&
\modelcell{2.65cm}{Zero-range BCS pairing + quadrupole--octupole shape mixing}
&
\modelcell{2.65cm}{No separate $g_R$; bare $M1$ operator}
\\ \hline

\modelcell{2.05cm}{RDFT + RAT-PRM~\cite{Wang2026CPB}}
&
\modelcell{2.65cm}{Self-consistent RDFT input + RAT-PRM}
&
\modelcell{2.05cm}{Input from self-consistent RDFT}
&
\modelcell{2.05cm}{Reflection-asymmetric rotor; good parity}
&
\modelcell{2.05cm}{particle--rotor coupling}
&
\modelcell{2.65cm}{Empirical pairing gap; no shape mixing}
&
\modelcell{2.65cm}{Adopted $g_R=Z/A$; fitted valence-neutron $g$ factor}
\\ \hline\hline

\end{tabular*}

\vspace{1mm}
\modelcell{0.98\textwidth}{\scriptsize
Abbreviations:
AMP, angular-momentum projection;
PP, parity projection;
PNP, particle-number projection;
BCS, Bardeen--Cooper--Schrieffer;
QO, quadrupole--octupole;
qp, quasiparticle.}
\end{table*}

\begin{figure*}[ht!]
\centering
\includegraphics[width=\textwidth]{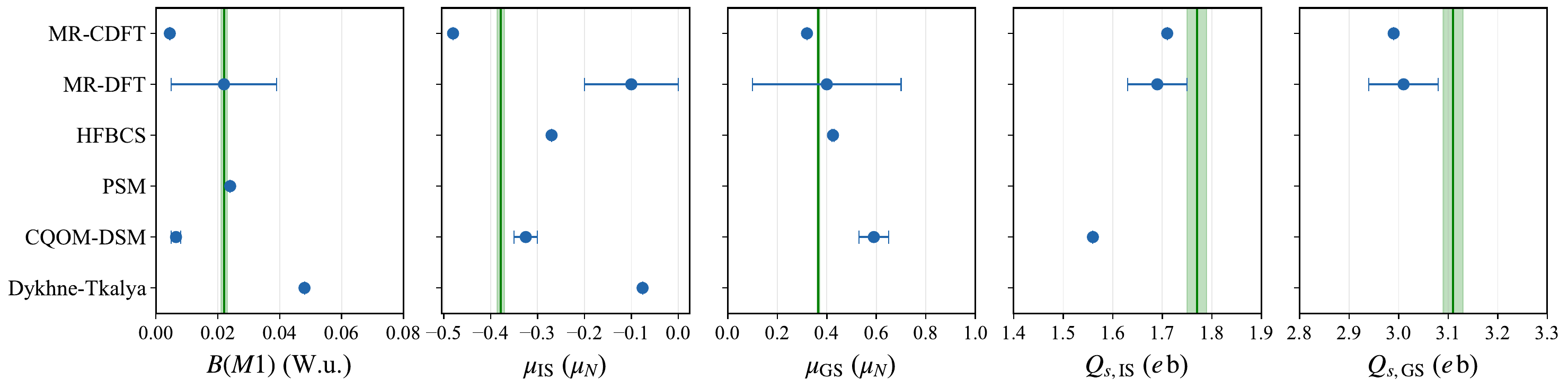}
\caption{ Comparison of theoretical predictions (blue circles with error bars) and experimental values (green vertical lines with shaded $1\sigma$ bands) for five electromagnetic observables of $^{229}$Th. From left to right: $M1$ transition probability $B(M1)$, isomeric magnetic moment $\mu_{\rm IS}$, ground-state magnetic moment $\mu_{\rm GS}$, isomeric spectroscopic electric quadrupole moment $Q_{\rm IS}^{s}$, and ground-state spectroscopic electric quadrupole moment $Q_{\rm GS}^{s}$. Experimental data are taken from Refs.~\cite{Tiedau2024physrevlettLaser,Yamaguchi2024natureLaser,Porsev2021PhysRevLettHyperfine}; theoretical values are taken from Refs.~\cite{Dykhne1998JetpLett.Matrix,Minkov2019Phys.Rev.Lett.Theoretical,Chen2025Phys.Lett.BPSM,Minkov2024Phys.Rev.CHFBCS,Restrepo2026DFT,zhou2025microscopic}. Models without a prediction for a given observable are omitted from the corresponding panel.}
\label{fig:forest}
\end{figure*}

Table~\ref{tab:theory-exp} collects the key nuclear structure observables predicted by the theoretical approaches reviewed above, together with the latest available data. The observables include the isomeric magnetic dipole transition probability $B(M1)$ and electric quadrupole transition probability $B(E2)$, the spectroscopic magnetic dipole moments $\mu_{\rm GS}$ and $\mu_{\rm IS}$, the spectroscopic quadrupole moments $Q_{\rm GS}^{s}$ and $Q_{\rm IS}^{s}$, and the difference between the mean-square charge radii of the isomeric and ground states, $\delta\langle r^2\rangle = \langle r^2\rangle_{\rm IS} - \langle r^2\rangle_{\rm GS}$.

\begin{table*}[ht!]
\centering
\caption{
Comparison of theoretical predictions with the latest available data for the key electromagnetic observables of the $^{229}$Th ground state ($5/2^+$, GS) and isomeric state ($3/2^+$, IS). 
$B(M1)$ and $B(E2)$ are in Weisskopf units (W.u.); magnetic dipole moments $\mu$ in units of the nuclear magneton $\mu_N$; spectroscopic electric quadrupole moments $Q_s$ in $e\,{\rm b}$; the difference between the mean-square charge radii of the isomeric and ground states, $\delta\langle r^2\rangle = \langle r^2\rangle_{\rm IS} - \langle r^2\rangle_{\rm GS}$, in fm$^2$. The $\mu_{\rm GS}$ and $Q_{\rm GS}^{s}$ values labeled ``Expt.'' are deduced from hyperfine-structure measurements via atomic-structure calculations~\cite{Porsev2021PhysRevLettHyperfine}.
}
\label{tab:theory-exp}

\renewcommand{\arraystretch}{1.4}

\begin{tabular}{lccccccc}
\hline\hline

& $B(M1)$ 
& $B(E2)$ 
& $\mu_{\rm GS}$ 
& $\mu_{\rm IS}$ 
& $Q_{\rm GS}^{s}$ 
& $Q_{\rm IS}^{s}$ 
& $\delta\langle r^2\rangle$ \\

& (W.u.) 
& (W.u.) 
& ($\mu_N$) 
& ($\mu_N$) 
  & ($e\,{\rm b}$)
  & ($e\,{\rm b}$)
& (fm$^2$) \\

\hline

Expt. 
& $0.022$~\cite{Tiedau2024physrevlettLaser} 
& --- 
& $0.366(6)$~\cite{Porsev2021PhysRevLettHyperfine} 
& $-0.378(8)$~\cite{Yamaguchi2024natureLaser} 
& $3.11(2)$~\cite{Porsev2021PhysRevLettHyperfine} 
& $1.77(2)$~\cite{Yamaguchi2024natureLaser} 
& $0.0097(26)$~\cite{Yamaguchi2024natureLaser} \\

\hline

Dykhne \& Tkalya 1998~\cite{Dykhne1998JetpLett.Matrix} 
& $0.048$ 
& --- 
& --- 
& $-0.076$ 
& --- 
& --- 
& --- \\

CQOM-DSM~\cite{Minkov2017Phys.Rev.Lett.Reduced,Minkov2019Phys.Rev.Lett.Theoretical} 
& $0.005$--$0.008$ 
& $20$--$30$ 
& $0.53$--$0.65$ 
& $-0.30$--$-0.35$ 
& --- 
& $1.56$ 
& --- \\

PSM~\cite{Chen2025Phys.Lett.BPSM} 
& $0.024$ 
& $8.7$--$11.0$ 
& --- 
& --- 
& --- 
& --- 
& --- \\

HFBCS~\cite{Minkov2024Phys.Rev.CHFBCS} 
& --- 
& --- 
& $0.42$--$0.43$ 
& $-0.27$ 
& --- 
& --- 
& --- \\

MR-DFT~\cite{Restrepo2026DFT} 
& $0.022(17)$ 
& --- 
& $0.4(3)$ 
& $-0.1(1)$ 
& $3.01(7)$ 
& $1.69(6)$ 
& --- \\

MR-CDFT~\cite{zhou2025microscopic}
& $0.0045$
& $5.6$
& $0.32$
& $-0.48$
& $2.99$
& $1.71$
& --- \\

\hline\hline
\end{tabular}
\end{table*}

Several trends emerge from Table~\ref{tab:theory-exp}; these comparisons are also visualized in Fig.~\ref{fig:forest}. The $B(M1)$ value provides a central benchmark for the nuclear-structure models. The early Nilsson-plus-Coriolis estimate of Dykhne and Tkalya~\cite{Dykhne1998JetpLett.Matrix} gives $B(M1) \approx 0.048$ W.u., about a factor of two above the experimental benchmark of $0.022$ W.u.~\cite{Tiedau2024physrevlettLaser}. The CQOM-DSM~\cite{Minkov2017Phys.Rev.Lett.Reduced,Minkov2019Phys.Rev.Lett.Theoretical},  including Coriolis mixing between the quadrupole-octupole motion of the core and the odd nucleon in a reflection-asymmetric deformed potential, predicts a smaller value of $0.005$--$0.008$ W.u. The PSM~\cite{Chen2025Phys.Lett.BPSM} calculation gives $B(M1) \approx 0.024$ W.u., close to the experimental benchmark, whereas the MR-DFT result~\cite{Restrepo2026DFT}, $B(M1)=0.022(17)$ W.u., is consistent within its quoted uncertainty. The MR-CDFT calculation of Zhou and Yao~\cite{zhou2025microscopic} gives a smaller value, $B(M1)=0.0045$ W.u., closer to the CQOM-DSM range than to the lifetime-inferred benchmark.   For the competing $E2$ branch of the isomeric transition, no experimental $B(E2)$ value is yet available; the CQOM-DSM predicts $20$--$30$ W.u., the PSM gives $8.7$--$11.0$ W.u., and the MR-CDFT calculation of Zhou and Yao gives $B(E2;3/2_1^+\rightarrow 5/2_1^+)\simeq 5.6$ W.u. from the interband transition systematics shown in their Fig.~21~\cite{zhou2025microscopic}. This spread indicates a model-dependent quadrupole strength. 

The isomeric magnetic moment $\mu_{\rm IS}$ is a stringent discriminator among models. The Dykhne--Tkalya estimate, $\mu_{\rm IS}=-0.076\,\mu_N$~\cite{Dykhne1998JetpLett.Matrix}, is far from the experimental value of $-0.378(8)\,\mu_N$~\cite{Yamaguchi2024natureLaser}. The CQOM-DSM result with a quenched rotational gyromagnetic factor $g_R$~\cite{Minkov2019Phys.Rev.Lett.Theoretical} lies close to experiment, whereas the HFBCS~\cite{Minkov2024Phys.Rev.CHFBCS} and MR-DFT~\cite{Restrepo2026DFT} calculations give smaller magnitudes, with the latter underestimating $|\mu_{\rm IS}|$ by a factor of $\sim 3$. By contrast, the MR-CDFT value of Zhou and Yao, $\mu_{\rm IS}=-0.48\,\mu_N$, somewhat overestimates the magnitude but has the correct sign and scale~\cite{zhou2025microscopic}.

The ground-state magnetic moment $\mu_{\rm GS}$ shows the opposite trend: the CQOM-DSM overestimates the experimental value of $0.366(6)\,\mu_N$ (CQOM-DSM: $0.53$--$0.65\,\mu_N$), whereas the self-consistent mean-field approaches give smaller values closer to experiment (HFBCS: $0.42$--$0.43\,\mu_N$; MR-DFT: $0.4(3)\,\mu_N$; MR-CDFT: $0.32\,\mu_N$). The PSM~\cite{Chen2025Phys.Lett.BPSM} does not report magnetic moments.

The spectroscopic quadrupole moments provide an independent benchmark for the models. The MR-DFT prediction $Q_{\rm GS}^{s}=3.01(7)\,e\,\mathrm{b}$ agrees reasonably well with the value $3.11(2)\,e\,\mathrm{b}$ deduced from hyperfine-structure measurements~\cite{Porsev2021PhysRevLettHyperfine}, indicating that the ground-state quadrupole deformation is captured. The MR-CDFT value $Q_{\rm GS}^{s}=2.99\,e\,\mathrm{b}$ reaches a similar level of agreement~\cite{zhou2025microscopic}. For the isomer, the CQOM-DSM prediction $Q_{\rm IS}^{s}=1.56\,e\,\mathrm{b}$ underestimates the experimental value of $1.77(2)\,e\,\mathrm{b}$~\cite{Yamaguchi2024natureLaser}, whereas the MR-DFT result $1.69(6)\,e\,\mathrm{b}$ and the MR-CDFT result $1.71\,e\,\mathrm{b}$ are consistent within the quoted uncertainty.

Beyond observable-by-observable comparisons, two cross-cutting patterns deserve emphasis. (i) The CQOM-DSM, which uses an adopted octupole deformation, overestimates $\mu_{\rm GS}$ but reproduces $\mu_{\rm IS}$ rather well when a quenched value of $g_R$ is used; the self-consistent EDF approaches (HFBCS, MR-DFT, and MR-CDFT) show the opposite imbalance, giving values closer to $\mu_{\rm GS}$ while underestimating $|\mu_{\rm IS}|$. The MR-CDFT result breaks this simple pattern: it slightly underestimates $\mu_{\rm GS}$ but overestimates $|\mu_{\rm IS}|$, while reproducing both quadrupole moments well. The persistence of this dual offset across methodologically distinct frameworks suggests that the $3/2[631]$ isomeric configuration is more sensitive than the $5/2[633]$ ground-state configuration to magnetic polarization and configuration-mixing effects, which are not yet captured uniformly across models. (ii) The MR-DFT calculation reproduces $B(M1)$ within its quoted regression uncertainty and gives spectroscopic quadrupole moments close to experiment, while still underestimating $|\mu_{\rm IS}|$. Since $B(M1)$ and $\mu_{\rm IS}$ involve the same $M1$ operator but probe off-diagonal and diagonal matrix elements, respectively, this pattern suggests that the transition matrix element entering $B(M1)$ is better constrained than the diagonal isomeric matrix element entering $\mu_{\rm IS}$. This may reflect the greater sensitivity of the diagonal moment to the detailed single-particle components and magnetic response of the isomeric wave function, quantities that are only indirectly constrained in the regression procedure. The smaller MR-CDFT value of $B(M1)$, despite its reasonable diagonal moments, reinforces that diagonal moments and transition matrix elements remain differently sensitive probes of the calculated wave functions.

None of the structure models discussed above currently reproduces the $\sim 8$ eV isomer energy as a genuine prediction. This limitation is not merely technical: the total binding energy of $^{229}$Th is approximately 1748 MeV~\cite{Wang2021ChinesePhys.CAME}, so the isomer energy corresponds to a relative scale of $8~\mathrm{eV}/1748~\mathrm{MeV}\sim 5\times 10^{-9}$. Although excitation energies are differences between nuclear states rather than absolute binding energies, this ratio illustrates the extreme cancellation required to reach the eV scale. Present nuclear Hamiltonians, whether phenomenological or derived from chiral effective field theory, are not constrained at this level for such a heavy deformed nucleus. Self-consistent mean-field calculations (HFBCS, MR-DFT, and MR-CDFT) reproduce the relevant near-Fermi orbital structure but predict splittings in the keV range; the PSM reduces the mean-field gap from $\sim 75$ keV through angular-momentum and parity projection together with configuration mixing, yet still does not reach the eV scale; and the CQOM-DSM avoids this difficulty by taking the experimental level energies as input. In current structure models, the observed isomer energy should therefore be regarded as an experimental constraint rather than as a benchmark already explained by nuclear-structure theory, while electromagnetic observables such as $B(M1)$ and the magnetic and quadrupole moments provide more meaningful benchmarks for the wave functions.

\section{Applications to Fundamental Physics and Quantum Technologies}\label{sec:newphysics}

The unusual nuclear structure and optical accessibility of the $^{229}\text{Th}$ isomer provide a compelling platform for precision metrology. This remarkably low energy scale arises from a cancellation among MeV-scale electromagnetic and strong interaction contributions, leaving a residual transition energy in the eV regime.  Since this energy is the small difference between two large, competing MeV-scale contributions, the transition frequency can lead to a large amplification of sensitivity to variations in fundamental interactions. This enhanced sensitivity, combined with the reduced sensitivity of nuclear degrees of freedom to many external perturbations due to electronic shielding, motivates diverse frontiers in fundamental and applied physics, such as searching for variations in fundamental constants \cite{Flambaum2006PhysRevLettEnhanced,He2007JPhysGEnhanced,He2008NuclPhysATemporal,Litvinova2009PhysRevCNuclear,Berengut2009PhysRevLettProposed, Flambaum2009PhysRevCEnhanced, Rellergert2010PhysRevLettConstraining, Uzan2011LivRevRelVarying, Fadeev2020PhysRevASensitivity, Fadeev2022PhysRevCEffects, Beeks2025NatCommunFineStructure, Uzan2025LivRevRelFundamental,Wang2025PhysRevASearch,Arakawa2026Probing,Caputo2025PhysRevCSensitivity}, probing dark matter and fifth forces \cite{Derevianko2014NatPhysHunting, Arvanitaki2015PhysRevDSearching, VanTilburg2015PhysRevLettSearch, Hees2016PhysRevLettSearching, Caputo2025PhysRevCSensitivity, Delaunay2025PhysRevDProbing, Fuchs2025PhysRevXSearching}, testing the equivalence principle \cite{Flambaum2016PhysRevLettEnhancing, Dzuba2025PhysRevAUsing, Peik2021QuantSciTechnolNuclear, Uzan2011LivRevRelVarying, Uzan2025LivRevRelFundamental}, and advancing relativistic geodesy \cite{Bondarescu2015EPJWebConfThePotential, Mehlstaubler2018RepProgPhysAtomic, Kolkowitz2016PhysRevDGravitational, Thirolf2019JPhysBThe229Thorium, Peik2021QuantSciTechnolNuclear} or nuclear quantum information \cite{Raeder2011JPhysBResonance, Tkalya2011PhysRevLettProposal, vonDerWense2018MeasTechTowards, Thirolf2019JPhysBThe229Thorium, Dzuba2025PhysRevAUsing, Perera2025PRHostDependent} processing. \red{The fundamental-physics discussion below is organized according to characteristic signal type: secular drifts of fundamental constants, oscillatory variations induced by ultralight dark matter, and static or anisotropic signals in tests of fundamental symmetries. Nuclear quantum technologies are treated separately as an application.}

\subsection{\texorpdfstring{\red{Secular Drifts of Fundamental Constants}}{Secular Drifts of Fundamental Constants}}

\label{subsec:alpha}
The most prominent manifestation of these sensitivity gains is found in the search for variations of the fine-structure constant, $\alpha$. The sensitivity of the nuclear transition frequency $\omega$ to variations in $\alpha$ is parameterized by the dimensionless enhancement factor $K_\alpha$, which scales the fractional frequency shift to the fractional change in $\alpha$ \cite{Flambaum2006PhysRevLettEnhanced}:
\begin{equation}
    \frac{\delta \omega}{\omega} = K_\alpha \frac{\delta \alpha}{\alpha}.
    \label{eq:enhancement_factor}
\end{equation}
In standard atomic optical clocks, transitions occur within the valence electronic shell, where the sensitivity arises principally from relativistic corrections. For these electronic transitions, the sensitivity coefficient $|K_\alpha|$ typically ranges from $ \approx 0.01$ in light ions (e.g., Al$^+$) to $\approx 6$ in heavy systems (e.g., Yb$^+$ $E3$ transition), generally remaining of order unity ($|K_\alpha| \sim 1$) \cite{Flambaum2006PhysRevLettEnhanced,Safronova2018Search}.

In contrast, the sensitivity of $^{229}$Th arises from the cancellation between the large Coulomb energy difference $\Delta V_C$ and the strong interaction energy difference $\Delta V_N$ between the isomer and the ground state. This enhancement can be understood through a simple scaling argument \cite{Flambaum2006PhysRevLettEnhanced}: assuming the strong interaction is independent of $\alpha$, the transition energy $\hbar \omega = \Delta V_C + \Delta V_N$ responds to variations in $\alpha$ primarily through the Coulomb term ($V_C \propto e^2 \propto \alpha$). Consequently, the enhancement factor simplifies to the ratio of the Coulomb energy difference to the net transition energy:
\begin{equation}
    K_\alpha = \frac{\delta\omega/\omega}{\delta \alpha / \alpha} \approx   \frac{\Delta V_C}{\hbar \omega}.
    \label{eq:K_scaling}
\end{equation}
Given that $\Delta V_C$ in heavy nuclei can be on the order of MeV ($10^6$ eV), while the transition energy is only $\sim 8$ eV, Eq.~(\ref{eq:K_scaling}) suggests a potentially enormous amplification factor, naively of order $K_\alpha \sim 10^5$. More detailed nuclear calculations and recent sensitivity analyses give values typically discussed at the $10^4$--$10^5$ level, but also emphasize that the precise coefficient is model dependent and can be affected by cancellations among nuclear contributions \cite{Flambaum2006PhysRevLettEnhanced,Berengut2009PhysRevLettProposed,Caputo2025PhysRevCSensitivity}. Even with this caveat, the expected enhancement exceeds that of most atomic systems by several orders of magnitude.

Currently, optical atomic-clock comparisons constrain the secular drift of the fine-structure constant to $\frac{1}{\alpha}\frac{d\alpha}{dt}=1.8(2.5)\times10^{-19}\,\mathrm{yr}^{-1}$ using repeated Yb$^+$ $E3/E2$ frequency-ratio measurements~\cite{PhysRevLett.130.253001}.
If the nuclear enhancement factor is indeed at the $10^4$ level, a $^{229}$Th nuclear clock could become competitive with existing atomic-clock bounds before reaching the same fractional frequency stability.  Complementary electron--nucleus-coupled schemes have also been proposed. In highly charged $^{229}$Th ions, HEB transitions can enhance the $\alpha$-variation sensitivity relative to the bare nuclear transition~\cite{Wang2025PhysRevASearch}. In Th~III, the EB process mainly provides enhanced routes for nuclear excitation and decay, whereas the proposed new-physics searches exploit the low-lying metastable electronic state, its weak M2 transition, and comparisons between electronic and nuclear clock frequencies~\cite{Dzuba2025PhysRevAUsing}. Recent microscopic nuclear structure calculations~\cite{zhou2025microscopic, zhou2025effects} also provide refined theoretical inputs essential for interpreting such precision measurements.

Beyond the electromagnetic sector, the $^{229}\text{Th}$ transition provides a highly sensitive probe for variations in the strong interaction, typically parameterized by the dimensionless ratio $X_q = m_q / \Lambda_{\text{QCD}}$ (where $m_q$ is the light quark mass and $\Lambda_{\text{QCD}}$ is the \red{quantum chromodynamics (QCD) scale}).
While electronic transitions in atomic clocks are largely decoupled from the hadronic sector, the nuclear transition frequency $\omega$ responds to variations in $X_q$ via a hadronic enhancement factor $K_q$:
\begin{equation}
\frac{\delta \omega}{\omega} = K_q \frac{\delta X_q}{X_q}.
\label{eq:K_q}
\end{equation}
Following the same logic, $K_q \approx \Delta V_N / \hbar \omega$. Since $\Delta V_N$ is also on the MeV scale, this yields an amplification of $K_q \sim 10^4 - 10^5$ \cite{Flambaum2006PhysRevLettEnhanced, Flambaum2009PhysRevCEnhanced}. Theoretical analyses within the Relativistic Mean Field framework highlight that a reliable determination of $K_q$ must account for the intricate correlation between strong and Coulomb forces, as the subtle polarization of the proton distribution due to neutron excitation provides a significant contribution to the sensitivity \cite{He2008NuclPhysATemporal}. 

At present, stringent constraints on variations of $X_q$ come from the Oklo natural nuclear reactor and big bang nucleosynthesis. In the Oklo analysis, the bound $|\delta X_q/X_q|<4\times10^{-9}$ over the past 1.8 billion years corresponds, under the assumption of linear time dependence, to $|\dot X_q/X_q|<2.2\times10^{-18}\,\mathrm{yr}^{-1}$~\cite{Flambaum2009PhysRevCEnhanced}; translating geological and cosmological constraints more generally into a per-year drift rate is model dependent~\cite{Uzan2025LivRevRelFundamental}.
The $^{229}\text{Th}$ nuclear clock could provide a high-precision laboratory platform to test these variations directly.   
By offering simultaneous enhanced sensitivity to both $\alpha$ and $X_q$, a $^{229}\text{Th}$ nuclear clock would provide a distinctive laboratory probe of correlated variations in electromagnetic and hadronic parameters, with direct relevance to beyond-Standard-Model scenarios such as ultralight dark matter.

\subsection{\texorpdfstring{\red{Oscillatory Signals from Ultralight Dark Matter}}{Oscillatory Signals from Ultralight Dark Matter}}
\label{subsec:darkmatter}

The $^{229}\text{Th}$ nuclear transition provides a particularly sensitive probe for \red{ultralight dark matter (ULDM)}. ULDM typically consists of hypothetical \red{light bosonic particles},  such as dilatons, moduli, or axion-like particles, with masses $m_{\text{DM}} \ll 1$ eV. Due to their extremely small masses and high number densities in the galactic halo, these particles exhibit high occupation numbers and behave as a classical, coherently oscillating background field, $\phi(t) = \phi_0 \cos(m_{\text{DM}} t)$, rather than as individual particles \cite{Arvanitaki2015PhysRevDSearching, Derevianko2014NatPhysHunting}. This field couples to Standard Model fields via non-gravitational interactions, inducing periodic modulations in fundamental constants such as the fine-structure constant $\alpha$, the light quark mass $m_q$, and the QCD energy scale $\Lambda_{\text{QCD}}$.

With a convention in which the scalar field is made dimensionless, for example $\varphi(t)=\sqrt{4\pi G_N}\,\phi_0\cos(m_{\rm DM}t)$, the resulting fractional frequency variation of the $^{229}\text{Th}$ transition can be written schematically as
\begin{equation}
  \frac{\delta \omega(t)}{\omega} = \left( K_\alpha d_e + K_q d_{\rm h} \right) \varphi(t),
  \label{eq:DM_oscillation_short}
\end{equation}
where $d_e$ denotes the electromagnetic coupling and $d_{\rm h}$ denotes the effective hadronic coupling combination relevant to the variation of $X_q=m_q/\Lambda_{\rm QCD}$, for example the combination of light-quark-mass and gluonic couplings that controls $\delta\ln X_q$ in a given convention. Conventions vary across the literature: the scalar-field normalization may be written explicitly or absorbed into effective coupling constants, and the gluonic and light-quark-mass couplings may be kept as separate parameters rather than combined into a single effective hadronic coupling.  While optical atomic clocks are mainly sensitive to photon- and electron-sector couplings,   the $^{229}\text{Th}$ system offers a particularly useful handle on hadronic-sector couplings. As discussed in Sec.~\ref{subsec:alpha}, the near-cancellation of large nuclear energy scales yields a hadronic enhancement factor $K_q$ of order $10^4$--$10^5$, which can substantially improve sensitivity to quark- and gluon-sector couplings in the relevant mass and coupling ranges~\cite{Fuchs2025PhysRevXSearching, Caputo2025PhysRevCSensitivity, Arakawa2026Probing}.

Recent experimental milestones have already begun to leverage this potential. Initial constraints on hadronic ULDM have been derived by analyzing the $^{229}\text{Th}$ nuclear lineshape through laser spectroscopy \cite{Fuchs2025PhysRevXSearching}. Most notably, a direct frequency comparison between the $^{229}\text{Th}$ transition and a strontium (Sr) optical clock has set strong limits on ULDM in the mass range of $10^{-21}\text{ eV} \lesssim m_{\text{DM}} \lesssim 10^{-19}\text{ eV}$ \cite{Arakawa2026Probing}. This measurement probed effective interaction scales above the Planck scale in that mass range, demonstrating the potential of the $^{229}\text{Th}$ nuclear clock for exploring dark-matter interactions in the nuclear sector \cite{Arakawa2026Probing}. The feedback-loop Th:CaF$_2$ clock further extends this program to continuous Th/Yb$^+$ clock operation, searching for oscillatory and drifting variations of the nuclear transition frequency on time scales from 20~s to about one day~\cite{DeCol2026Feedback}.  

\subsection{\texorpdfstring{\red{Static and Anisotropic Signals in Fundamental-Symmetry Tests}}{Static and Anisotropic Signals in Fundamental-Symmetry Tests}}
\label{subsec:symmetries}

The octupole-correlated nuclear structure \red{and low-energy transition} of $^{229}$Th \red{make} it a useful setting for testing fundamental symmetries, specifically parity ($P$) and time-reversal ($T$) invariance, as well as local Lorentz invariance \red{violation} (LLIV). \red{For $P$- and $T$-violation,} the primary driver for sensitivity enhancement in this system is the collective nature of the nucleus, associated with strong octupole correlations (a pear-shaped geometry).


In the sector of $P$- and $T$-violation, such deformation leads to the emergence of nearly degenerate parity doublets, i.e., pairs of nuclear states with opposite parity and small energy separation $\Delta E_{\pm} = E_- - E_+$. According to the standard framework, the nuclear Schiff moment $S$, which acts as the primary source of atomic electric dipole moments in heavy atoms, is generated by the mixing of these opposite-parity states via $P,T$-odd nuclear forces. In nuclei with octupole features, this mixing is dominated by the closest-lying doublet, yielding an enhanced collective moment \cite{Auerbach1996PhysRevLettCollective, Flambaum2019PhysRevCEnhanced}:
\begin{equation}
S \propto  \frac{\langle \Psi_- | \hat{H}_{P,T} | \Psi_+ \rangle \langle \Psi_+ | \hat{S}_{\rm coll} | \Psi_- \rangle}{\Delta E_{\pm}},
\label{eq:Schiff_enhancement}
\end{equation}
where $\hat{H}_{P,T}$ represents the $P$- and $T$-violating interaction and $\hat{S}_{\rm coll}$ is the collective Schiff operator. As discussed in Refs.~\cite{Auerbach1996PhysRevLettCollective, Flambaum2019PhysRevCEnhanced}, this splitting $\Delta E_{\pm}$ is naturally small in $^{229}$Th compared to spherical nuclei. Specifically, the mixing involves the $5/2 [633]$ ground state and the $5/2^-$ band head at approximately $146$ keV. A recent MR-CDFT calculation further predicts an unobserved $5/2^-$ parity partner of the ground-state band at about 15 keV; if such a state exists, the smaller parity-doublet splitting would enhance the Schiff moment by approximately one order of magnitude~\cite{zhou2025microscopic}. While the $\sim 8$ eV isomer is of positive parity ($3/2^+$), its existence provides an optical handle to interrogate the nucleus via precision spectroscopy. The combination of low-lying negative parity excitations and the collective octupole moment results in an enhancement of the $T$-violating atomic electric dipole moment by factors of $10^2$--$10^3$ compared to $^{199}$Hg \cite{Auerbach1996PhysRevLettCollective, Flambaum2019PhysRevCEnhanced}. When embedded in polar molecules such as $^{229}$ThO or $^{229}$ThF$^+$, the large internal molecular electric fields further amplify these effects, offering a sensitive probe for CP-violation beyond the Standard Model \cite{Flambaum2019PhysRevCEnhanced}.

The $^{229}$Th system is also useful for LLIV and Einstein-equivalence-principle-type clock comparisons, but these tests should be distinguished from the static Schiff-moment searches discussed above. In the Standard-Model Extension framework, these effects are described by kinetic-energy and tensor operators rather than by ordinary static electromagnetic moments. Nuclear deformation and quadrupole collectivity provide useful structural guidance, but the relevant LLIV observable is not simply the ordinary spectroscopic electric quadrupole moment. The enhanced fractional sensitivity arises because the transition energy $\omega \sim 8$~eV is the residual of a strong cancellation among MeV-scale nuclear contributions. Consequently, any small LLIV- or Einstein-equivalence-principle-violating (EEPV) differential shift $\delta \omega_{\rm LLIV/EEPV}$ becomes fractionally amplified by roughly the ratio of a MeV-scale matrix-element difference to the eV-scale transition energy, i.e., by a factor of order $10^5$~\cite{Flambaum2016PhysRevLettEnhancing}:
\begin{equation}
\frac{\delta \omega_{\rm LLIV/EEPV}}{\omega} =
\frac{\Delta \langle \hat{H}_{\rm LLIV/EEPV} \rangle}{\omega},
\end{equation}
where $\Delta \langle \hat{H}_{\rm LLIV/EEPV} \rangle$ represents the differential energy shift between the isomer and  the ground state. Comparing the $^{229}$Th nuclear transition with state-of-the-art electronic clocks, or exploiting the combined electronic and nuclear transitions in Th III, has therefore been proposed to improve tests of LLIV and EEPV~\cite{Flambaum2016PhysRevLettEnhancing, Dzuba2025PhysRevAUsing}.

A separate weak-interaction observable is the nuclear weak quadrupole moment. The favorable electronic structure of the Th III ion has been proposed as a route to access this quantity through enhanced parity-violating mixing, providing a complementary probe of the neutron-distribution quadrupole moment rather than the same observable as the LLIV or Einstein-equivalence-principle clock comparison~\cite{Dzuba2025PhysRevAUsing}.

\subsection{Nuclear Quantum Information and Technology}
\label{subsec:quantumtech}

The $^{229}\text{Th}$ system is not only a powerful tool for fundamental tests but also offers a potential platform for nuclear quantum information processing, owing to the intrinsic isolation of nuclear degrees of freedom from environmental perturbations. The primary figure of merit for its performance as a quantum bit or a frequency standard is the quality factor $Q$, defined as the ratio of the transition frequency $\nu$ to the natural linewidth $\Delta \nu$:
\begin{equation}
Q = \frac{\nu}{\Delta \nu} = 2\pi \nu \tau,
\label{eq:Q_factor}
\end{equation}
where $\tau$ is the radiative lifetime of the excited state. For the $^{229}$Th isomer, the transition energy is $E \approx 8.36$~eV ($\nu \approx 2.02$~PHz). Given that the theoretical radiative lifetime of the magnetic dipole (M1) transition is on the order of $\tau \sim 10^3$--$10^4$~s \cite{Minkov2019Phys.Rev.Lett.Theoretical}, Eq.~(\ref{eq:Q_factor}) yields a theoretical quality factor of $Q \approx 10^{19}$--$10^{20}$. This is approximately three orders of magnitude higher than the best atomic clock transitions (typically $Q \sim 10^{16}$--$10^{17}$), implying potential for unprecedented timing precision and coherence times.

This potentially long coherence arises from the nuclear shielding effect. Unlike valence electrons in atomic qubits, nuclear spin states are spatially confined within the femtometer-scale nucleus ($R_{\rm nuc}\sim r_0A^{1/3}\simeq 7~{\rm fm}$) and are shielded from external electromagnetic perturbations by the surrounding electron shell. The sensitivity to magnetic noise, for instance, is intrinsically suppressed by the ratio of the nuclear magneton $\mu_N$ to the Bohr magneton $\mu_B$:
\begin{equation}
\eta_B \sim \frac{\mu_N}{\mu_B} \approx \frac{1}{1836} \approx 5 \times 10^{-4}.
\end{equation}
Beyond the metrological applications discussed earlier, this intrinsic noise decoupling suggests that $^{229}\text{Th}$ may be useful for nuclear qubits and long-lived quantum memories. Unlike conventional electronic qubits, these spatially confined nuclear states are less exposed to magnetic dephasing. Leveraging the coherent control schemes developed for nuclear-clock operation, this robustness can be mapped onto two complementary quantum hardware architectures. On the one hand, individually trapped $^{229}\text{Th}$ ions can be interfaced with mature ion-trap quantum computing frameworks, offering a controlled environment for exploring nuclear-state logic. On the other hand, doping $^{229}\text{Th}$ into wide-bandgap hosts like $\text{CaF}_2$ presents a pathway toward dense ensemble registers \cite{Perera2025PRHostDependent}. Operating under cryogenic conditions to preserve optical coherence, such solid-state ensembles may serve as scalable quantum memories or nodes within future quantum networks.

 \section{Summary and Outlook}\label{SP}

In this review, we have discussed the $^{229}$Th nuclear isomer from three complementary perspectives: experimental spectroscopy and clock development, nuclear structure theory, and applications to fundamental physics and quantum technologies.

On the experimental side, the field has moved from indirect energy inference to direct laser control of the nuclear transition. Key milestones include the identification of the isomer through internal-conversion electrons~\cite{VonDerWense2016NatureDirecta}, laser-spectroscopic determination of nuclear moments in trapped ions~\cite{Thielking2018NatureLasera,Yamaguchi2024natureLaser}, resonant laser excitation in wide-bandgap crystals and thin films~\cite{Tiedau2024physrevlettLaser,Elwell2024physrevlettLaser,Zhang2024NatureThF4}, absolute frequency comparison with an optical atomic clock~\cite{Zhang2024natureFrequency}, and independent demonstrations of feedback-loop operation of solid-state Th:CaF$_2$ nuclear clocks~\cite{DeCol2026Feedback,Huang2026NuclearClock}. These advances have fixed the transition energy at the $\sim 8$ eV scale with laser-spectroscopic precision and have established two main clock directions: trapped-ion systems, which target high systematic accuracy, and solid-state systems, which exploit large ensembles for high stability. Highly charged ions and internal-conversion-based thin-film systems provide complementary routes for studying nuclear-electronic coupling, active state manipulation, and compact readout.

On the theoretical side, the central structure problem is the near-degeneracy of the $5/2^+[633]$ and $3/2^+[631]$ neutron Nilsson configurations. Coriolis mixing explains why the $\Delta K=1$ $M1$ transition, which is strongly suppressed in the asymptotic Nilsson limit, can occur, but a reflection-symmetric Nilsson-plus-Coriolis description fails to reproduce the measured isomeric magnetic moment. The CQOM-DSM framework showed that octupole deformation can strongly modify the $M1$ matrix elements and bring $\mu_{\rm IS}$ close to experiment after collective $g_R$ quenching~\cite{Minkov2017Phys.Rev.Lett.Reduced,Minkov2019Phys.Rev.Lett.Theoretical}. More recent microscopic calculations, including the PSM~\cite{Chen2025Phys.Lett.BPSM}, Skyrme HFBCS~\cite{Minkov2024Phys.Rev.CHFBCS}, Skyrme MR-DFT~\cite{Restrepo2026DFT}, MR-CDFT~\cite{zhou2025microscopic}, and RDFT combined with a particle-rotor model~\cite{Wang2026CPB}, all point to octupole correlations as an important structural ingredient. They nevertheless differ in detail: PSM and Skyrme MR-DFT give $B(M1)$ values close to the lifetime-inferred benchmark, CQOM-DSM and MR-CDFT predict smaller $M1$ transition probabilities, and the description of $\mu_{\rm IS}$ remains model dependent. Since the eV-scale excitation energy has not yet been reproduced with the required accuracy, electromagnetic moments, transition probabilities, charge-radius changes, and deformation observables remain especially useful benchmarks for testing the calculated wave functions.

On the fundamental-physics side, the reduction of MeV-scale nuclear contributions to an $\sim 8$ eV transition can enhance the response of the clock frequency to variations of both the fine-structure constant $\alpha$ and hadronic parameters such as $X_q=m_q/\Lambda_{\rm QCD}$. This makes a $^{229}$Th nuclear clock a distinctive laboratory probe of correlated variations in electromagnetic and hadronic sectors, with particular relevance to ultralight dark matter searches~\cite{Fuchs2025PhysRevXSearching,Arakawa2026Probing,Caputo2025PhysRevCSensitivity}. The same nucleus also offers complementary sensitivity to several distinct symmetry and new-physics channels: $P$- and $T$-violating interactions through octupole-enhanced Schiff moments, possible violations of local Lorentz invariance and the Einstein equivalence principle through clock-frequency comparisons, and weak-interaction observables such as the nuclear weak quadrupole moment in suitable electronic systems~\cite{Flambaum2019PhysRevCEnhanced,Flambaum2016PhysRevLettEnhancing,Dzuba2025PhysRevAUsing}. It also motivates potential applications in nuclear quantum information based on the long radiative lifetime and reduced sensitivity of nuclear degrees of freedom to environmental perturbations.

Looking forward, a central question is not only how to build better $^{229}$Th clocks, but also why such an exceptionally low nuclear energy scale occurs in the first place. The $\sim 8$ eV splitting reflects a cancellation between MeV-scale electromagnetic and strong-interaction contributions; whether this cancellation is purely accidental or points to a deeper structural mechanism or hidden symmetry remains an open problem. Identifying any systematic pattern or symmetry behind such cancellations would also guide the search for other low-energy nuclear transitions suitable for clock applications. Another important direction is to move beyond the present separation between nuclear-structure calculations and atomic, ionic, or solid-state modeling. In most current treatments, these problems are handled in separate stages: nuclear-structure calculations provide the isomer energy, electromagnetic moments, transition matrix elements, charge-radius changes, and deformation properties, whereas atomic, ionic, or material calculations describe excitation, detection, frequency shifts, broadening, and quenching in a given platform. Current microscopic treatments already provide essential nuclear and electronic inputs, but the coupling between them is often introduced through platform-specific perturbative corrections. Because the $\sim 8$ eV nuclear splitting lies on an electronic energy scale, developing a unified compound-system Hamiltonian, or an equivalent self-consistent framework, remains an important theoretical challenge.
Progress will therefore depend on cross-calibration among precision spectroscopy, controlled ion and solid-state platforms, microscopic nuclear-structure theory, and atomic, molecular, and material modeling. In this sense, $^{229}$Th is becoming more than a newly accessible nuclear resonance: it is a quantitatively testable interface between nuclear structure, precision metrology, searches for physics beyond the Standard Model, and emerging nuclear quantum technologies.
 
\begin{acknowledgments}
 
Helpful discussion with H. Sagawa, W. Wu, Y. M. Jiang, and Y. D. Wang are highly appreciated. This work is supported by the Strategic Priority Research Program of the Chinese Academy of Sciences (Grant Nos. XDB0920000, \red{XDB1550100,} and XDB34010100), the National Natural Science Foundation of China (Grant Nos. 12447101, 12347139, 12375118, 12435008, and W2412043), and the National Key R\&D Program of China (Grant Nos. 2023YFA1606500 and 2024YFE0109800). 
The results described in this paper are obtained on the High-performance Computing Cluster of ITP-CAS and the ScGrid of the Supercomputing Center, Computer Network Information Center of Chinese Academy of Sciences. 
\end{acknowledgments}


\bibliography{th229-xlu.bib}
\end{document}